\documentclass[runningheads]{llncs} 
\usepackage{lscape}
\usepackage{url}
\usepackage{longtable}
\usepackage{float}
\usepackage{multirow}
\usepackage{amsmath}
\usepackage{graphicx}
\usepackage{subfigure}
\usepackage{amssymb}
\usepackage{lineno}
\usepackage{booktabs}
\usepackage{textcomp}

\begin{document}

\begin{frontmatter}
\title{Structure and Dynamic of Global Population Migration Network\thanks{Supported by the Chinese National Natural Science Foundation (71701018, 61673070); Humanities and Social Sciences Foundation of Ministry of Education of China (20YJAZH010); the National Social Sciences Fund, China (14BSH024).}
}
\author{Wensha Gou\inst{1}, Siyu Huang\inst{1}, Qinghua Chen\inst{1,2,3}\\ Jiawei Chen\inst{1} and Xiaomeng Li\inst{1}*}
\date{}

\institute{School of Systems Science, Beijing Normal University, Beijing, China \\ \and New England Complex Systems Institute, Cambridge, MA, USA\\ 
\and Department of Chemistry, Brandeis University, Waltham, MA, USA\\
*corresponding author: \email{lixiaomeng@bnu.edu.cn}}

\authorrunning{W. Gou et al.}

\maketitle    

\begin{abstract}

Cross-border migration brings economic and cultural impacts to the origin and destination, and is also a key to reflect the international relations of related countries. In fact, the migration relationships of countries are complex and multilateral, but most traditional migration models are bilateral. Network theories could provide a better description of global migration to show the structure and statistical characteristics more clearly. Based on the estimated migration data and disparity filter algorithm, the networks describing the global multilateral migration relationships has been extracted among 200 countries over fifty years. The results show that the global migration networks during 1960-2015 exhibit a clustering and disassortative feature, implying globalized and multipolarized changes of migration during these years. The networks were embed into a Poincar\'e disk, yielding a typical and hierarchical ``core-periphery" structure which, associated with angular density distribution, has been used to describe the ``multi-centering'' trend since 1990s. Analysis on correlation and evolution of communities indicates the stability of most communities yet some structural changes still exist since 1990s, which reflect that the important historical events are contributable to regional and even global migration patterns.


\end{abstract}

\keywords{global migration network \and  disassortativity \and multipolarization \and hyperbolic geometry \and complexity.}

\end{frontmatter}

\clearpage

\section{Introduction}

In the tide of globalization, the scale and diversity of international migration are substantially increasing~\cite{abel2014quantifying,Li2016Characterizing}. In 2015, approximately 244 million people, or 3.3\% of the world's population, lived in a country other than their birthplace~\cite{BeaucheminINTERNATIONAL2016,abel2019bilateral}, and this value is forecasted to double by 2050~\cite{RodrigoCommunity2016}. Population migration could bring important effects on both importing and exporting countries~\cite{educationalmigra2006,Beine2001Brain,Frederic2012}, and some scholars have used quantitative models to analyze the influencing factors, evolution patterns and trends of global population migration~\cite{eumigra2015,Li2016Characterizing}.

Some early studies researched the mechanism of population migration, such as the conventional gravity model~\cite{Ravenstein1884The,ZipfThe1946,PootGravity2016}, the random utility maximization (RUM) model~\cite{HansonIncome2011,Peri2013The}, and the self selection method~\cite{Roy1951,Borjas1987Self,Borjas1999Chapter}. However, they most focus on the bilateral migration flow and relations between two countries. In reality, potential migrants usually face multiple optional destinations at the same time, and they have to make a decision after comparing the advantages of all possible choices. In some cases, these decisions even could be random or probabilistic~\cite{Li2016Characterizing}. So the existing methods based on bilateral relationship will bring the problem of information loss and distortion, which has promoted the development of multilateral models and theories. Subsequent scholars put forward the definition of the multilateral migration barrier and introduced the structural gravity model~\cite{C2009Diasporas,Burger2009gravity,James2011The} and multilateral probability model~\cite{Li2016Characterizing,QinghuaAnalyzing2019,LiAnalyzing2020}. Such improvements from bilateral to multilateral analysis are meaningful yet still insufficient, because the indefinite definition of multilateral barriers and coupling parameters hinder the following quantitative work, and bring some controversy in estimation~\cite{BergstrandGravity2013}. Some improvement in the methods are still needed.

The asymmetric and multilateral flow data shows the complexity of global migration system, and it needs a systematic model to describe the individual choice based on multilateral relationship, where the influences of other countries should not be ignored when discussing the migration flow between any two countries~\cite{Li2016Characterizing,LiAnalyzing2020}. Complex network is a powerful framework to understand the multilateral relationships in real world, where nodes are world countries (or other elements) and links represent interaction channels between countries. And it can show the overall structure and statistical characteristics of the system more clearly, such as in biological systems~\cite{multilayer2017}, cortical circuits~\cite{Cell-type-specific2017}, geographic maps~\cite{geonetcomp2015} and so on. In recent years, some researchers have used the complex network method to study international migration and have achieved some preliminary results~\cite{RodrigoCommunity2016}. 

The statistical characteristics of the network, such as degree correlation, clustering coefficient, connectivity~\cite{M2003The}, could describe the preference of migrants when selecting the destination, and also indicate the global/local connectivity and topology structure of the migration network~\cite{Ridolfi2013Global,porat2016global,boccaletti2006complex}. Fagiolo and Mastrorillo were the first scholars to study migration from a complex-network perspective, and based on analyzing the statistical characteristics, they indicated that the global migration network was organized with a small-world binary pattern displaying the characteristics of disassortativity and high clustering~\cite{International-MigrationNetworke}. This finding was later certificated by other scholars~\cite{Ridolfi2013Global,porat2016global}. Furthermore, the identification of communities could analyze the hierarchical structure, the relationships and similarities of different countries~\cite{lancichinetti2010characterizing}. Porat and Benguigui analyzed the degree distribution and connectivity of global migration networks and classified 145 destination countries into three classes~\cite{IdanPorat2015}. Some scholars decomposed the world migration network into communities and analyzed its structure evolution~\cite{RodrigoCommunity2016}, along with the glocalization, polarization and globalization of the network~\cite{Valentin2016}. In addition, it can also help conventional models to describe multilateral relations more scientifically, as Tranos analyzed the topology of a migration network, and proposed the pull and push factors behind international migration flows between OECD countries with the network method and gravity model~\cite{Nijkamp2014International}. This paper comprehensively analyzes the statistical characteristics, topological structure and evolution trend of the global migration networks, composed of 200 countries/regions, with an evolution time greater than 50 years.

Besides, in recent years, scholars have found hyperbolic features in some real-world networks~\cite{Michele2015Hyperbolicity,SerranoThe2016}. And here we try to study the geometric features of population migration networks. In addition to proposing the hyperbolic characteristics of the population migration network, the geometric configuration is also helpful for intuitively analyzing the regional and global structures of the whole system.

This paper is organized as follows: Section 2 introduces the data source and method to extract the backbone network of global bilateral migration, which is called GMN (Global Migration Network) in following sections. Section 3 analyzes the skeletal construction and community dynamics of GMNs, including the changes in network statistical characteristics during 1960 to 2015, and the structure evolution. The results confirm that the GMN is a disassortative network with high clustering coefficient, exhibiting globalized and multipolarized changes during 1960-2015. Additionally, the network represents hyperbolic and hierarchical characteristics by embedding the countries on a Poincar\'e disk. Section 4 provides the conclusions and discussion.

\section{Materials and Methods}

\subsection{Data source}

The analysis requires the data of bilateral migration flow between countries. However, from the perspective of statistics, authoritative institutions generally only provide data on the composition of immigrants (immigration stock data), such as the `UN Global Migration database'~\cite{un.org.migration} and `World Bank Global Bilateral Migration database'~\cite{worldbank.org}, which cover most of the countries in the world. Some existing global migration networks are directly based on the immigrant stock data, which could represent past flow quantities~\cite{RodrigoCommunity2016,Ridolfi2013Global,International-MigrationNetworke}. 

In addition, there are three common methods to estimate the bilateral migration flows based on the immigrant stock data published by the World Bank or United Nations~\cite{abel2019bilateral}: 1. use the differences in successive bilateral stocks to estimate the corresponding migration flows~\cite{porat2016global,IdanPorat2015,Li2016Characterizing}; 2. approximate the migration flow rates, which are then multiplied by additional data to obtain the estimated global migration flows ~\cite{Dennett2016Estimating}; and 3. frame the changes in migrant stocks as the residuals in a global demographic account~\cite{abel2014quantifying,abel2018estimates}. Among the literature, the third method, called ``demographic accounting”, could estimate migration flows to match increases or decreases in the reported bilateral stocks with births and deaths during the period. Some scholars consider ``demographic accounting” with a Pseudo-Bayesian method as the most effective estimating method~\cite{abel2019bilateral,azose2019estimation}. And this paper uses the third method provided by Abel based on 200 countries/regions during 1960-2015~\cite{GuyAbelDatabase}. 

To reduce the impact of contingency, we separate the data into 6 periods: 1960-1969, 1970-1979, 1980-1989, 1990-1999, 2000-2009 and 2010-2015. Since there are only five complete years of data after 2010, we have doubled the estimated flow data of 2010-2015 to match other periods.

\subsection{Global Migration Network (GMN)}

From the data in the previous section, we construct an undirected complex network based on estimated bilateral migration flows for each period. Here, the nodes are the countries/regions, and the connection between nodes depends on whether there are migrant flows between them. The weight of the edge indicates the volume of migrants, which is the sum of emigrant and immigrant flows. 

Global migration is a complex system with complicated microstructure and evolutionary characteristics~\cite{International-MigrationNetworke,IdanPorat2015,porat2016global}. The backbone of the network offers a perspective where the structural characteristics of the network are more prominent. There are many ways to extract the backbone network, we apply a method called disparity filter algorithm~\cite{Serrano2009} (with details in Appendix \ref{appendix:Disparity filter algorithm}). 

\begin{figure}[htp]
    \centering
    \subfigure[]{
    \includegraphics[width=0.45\textwidth]{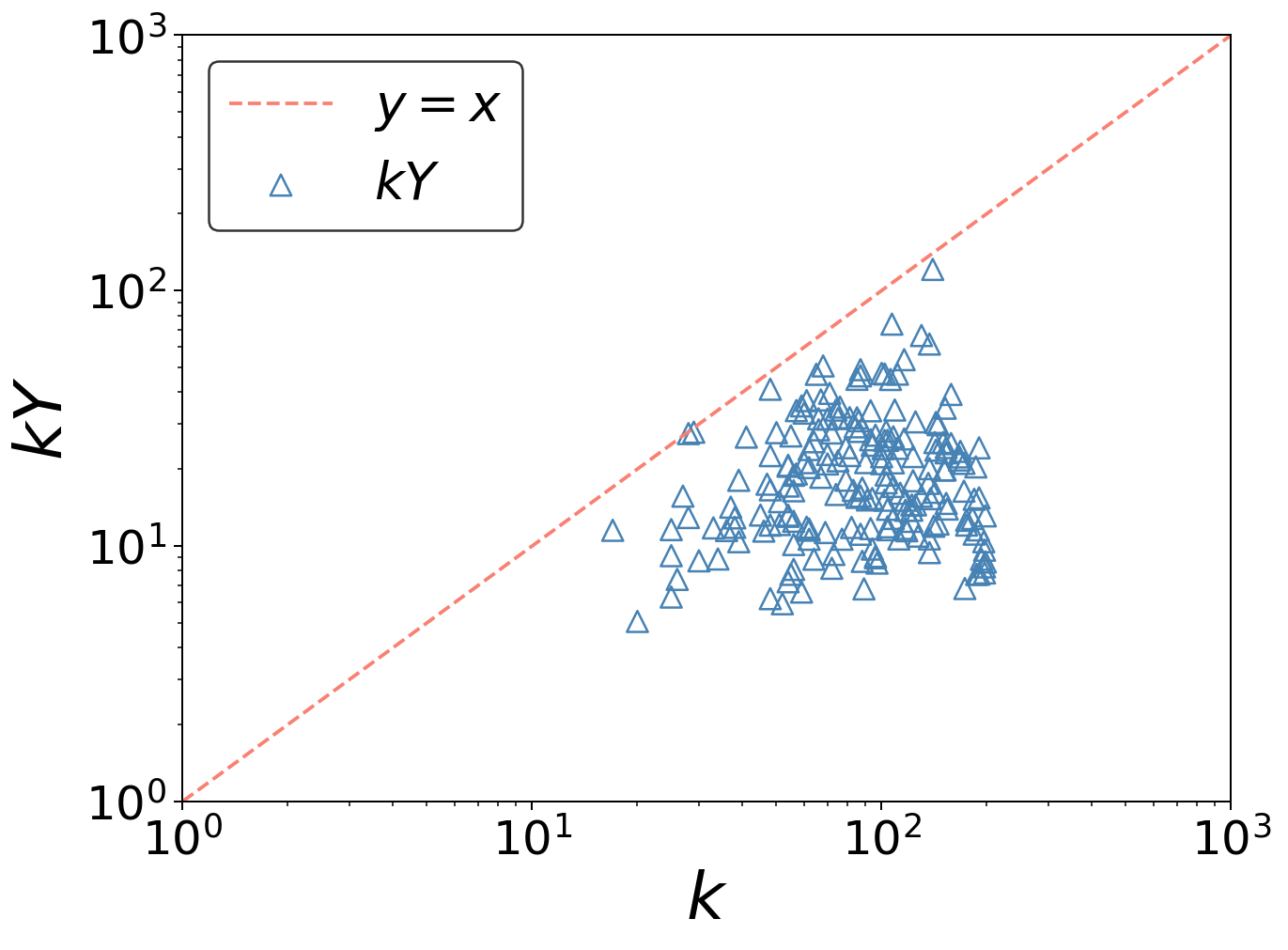}
}
    \subfigure[]{
    \includegraphics[width=0.45\textwidth]{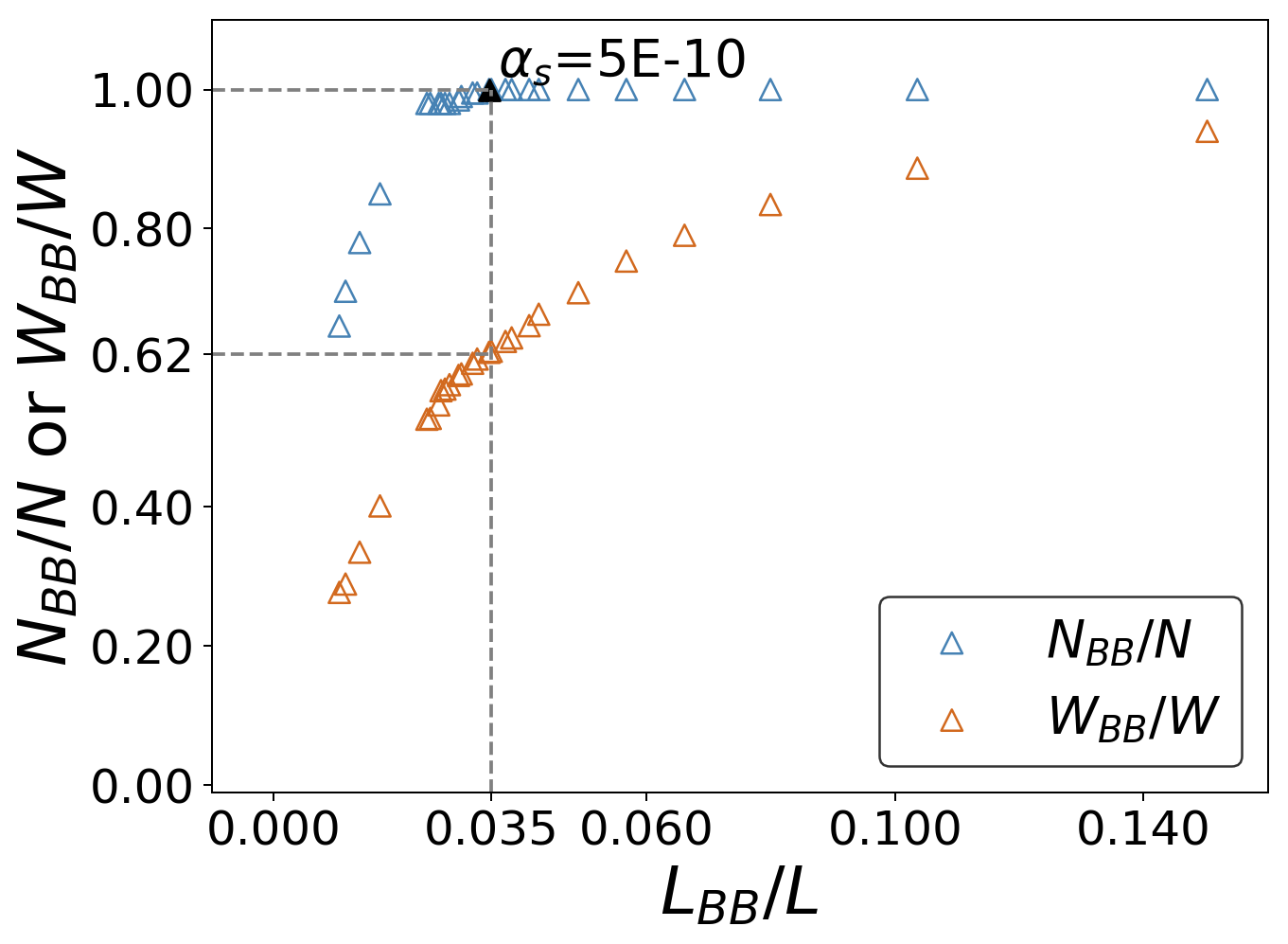}
}
    \caption{\textbf{(a)} Local inhomogeneity levels during the period of 2010-2015. \textbf{(b)} $N_{BB}/N$ and $W_{BB}/W$ vs. $L_{BB}/L$ in the period of 2010-2015. The position with a black triangle is the optimized extracting parameter $\alpha_s$ that we selected, retaining 3.47\% (385) of the edges and 62.22\% (5.95E+7) of the weight. }
    \label{fig:backbone extracting}
\end{figure} 

We first assess the effect of inhomogeneity at the local level; for each country $i$ with $k$ migration routes, we calculate the Herfindahl-Hirschman index $Y_i(k)=\sum_j(\frac{w_{i,j}}{\sum_jw_{i,j}})^2$ ~\cite{Serrano2009,SerranoThe2016} (in Appendix \ref{appendix:Disparity filter algorithm}). $w_{i,j}$ is the weight of the edge connecting $i$ and $j$. The local heterogeneity in the distribution of migration reveals that not all migration channels are equally significant (in Fig. \ref{fig:backbone extracting}(a), most blue triangles are below and near the red line of $y=x$, i.e., $kY<k$), and thus, the disparity filter can be applied to select only migration channels that are significant to at least one of the countries at the end of the channel.

\begin{figure}
    \centering
    \subfigure[1960-1969]{
    \includegraphics[width=1.1\textwidth]{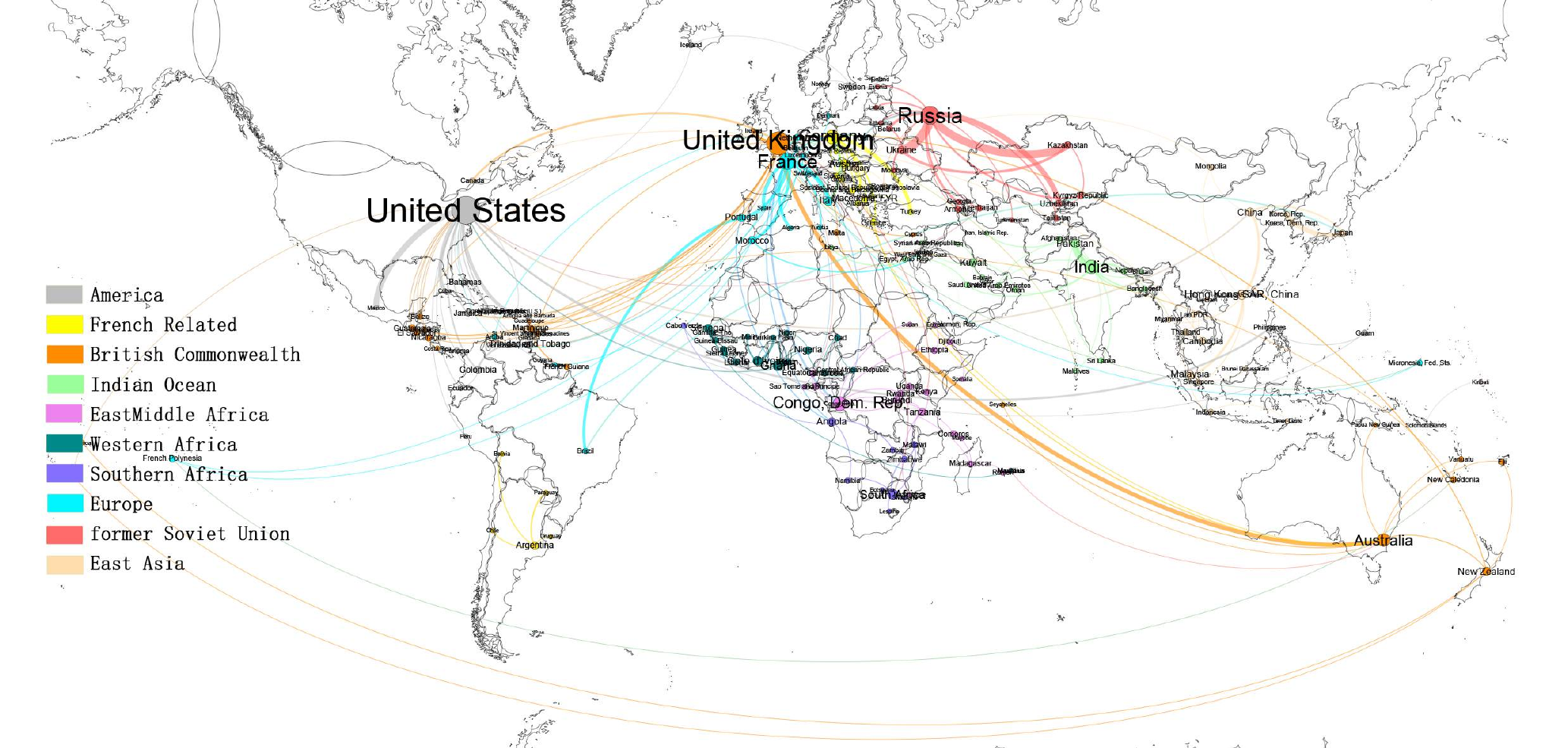}
}
    \subfigure[2010-2015]{
    \includegraphics[width=1.1\textwidth]{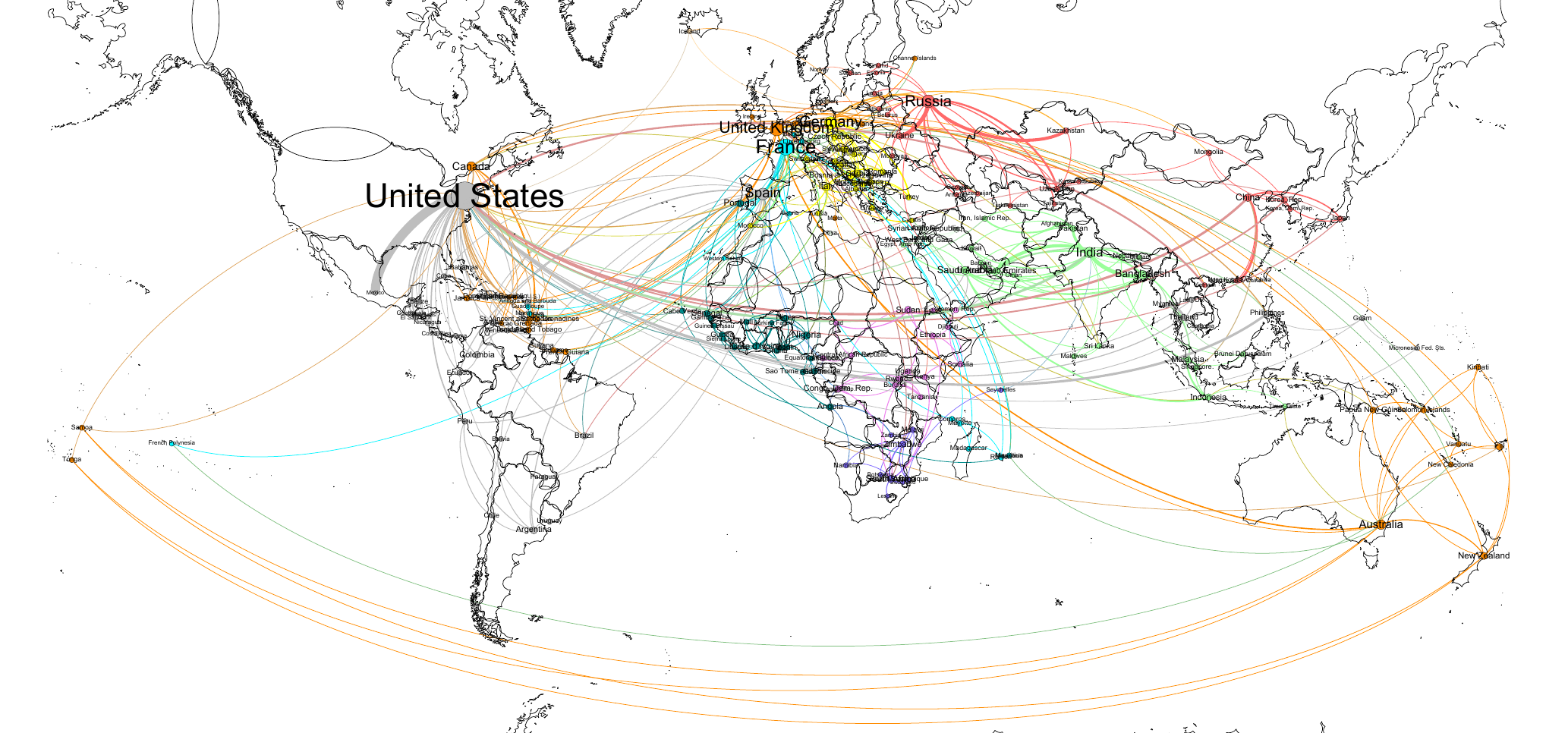}
}
     \caption{Backbone of global migration network.}
    \label{fig:backbone}
\end{figure}

$N_{BB}$, $L_{BB}$ and $W_{BB}$ are the number of nodes, number of links and total weights in the backbone network, respectively; while $N$, $L$ and $W$ are those in the original flow network. As significance level $\alpha$ changes from 0 to 0.1, the fraction of remaining nodes $N_{BB}/N$ and the fraction of remaining weights $W_{BB}/W$ gradually decrease, and the absolute value of the decreasing slope becomes increasingly large. To keep more countries, more weights and fewer links in the backbone network, we choose a position $\alpha_s$ (in Fig. \ref{fig:backbone extracting}(b)) which can minimize the remaining links on the premise of remaining all nodes to extract the backbone network. The extracted backbone network, which is called the global migration network (GMN) in the following sections, is shown in Fig. \ref{fig:backbone}.  The color of the node indicates the community of the country/region, which is consistent with Sections \ref{sec:communities} and \ref{section:structure evolution}. The results show that the GMN became denser in the 2010s.

\subsection{Hyperbolic geometry and embedding methods}

In a hyperbolic disk, the radius increases exponentially, and the distance between two points depends not only on the length of the line connecting the two points, but also on the angle difference. The hyperbolic space can capture the centrality and hierarchy of the network easily, and some scholars have proposed that many real-world networks exhibit natural hyperbolic geometry~\cite{BoguHyperbolic2010,SerranoThe2016}, ranging from biology to economics, finance and trade~\cite{Caldarelli2007Scale,Barthelemy2000,BoguHyperbolic2010,Bashan2013The,Michele2015Hyperbolicity,SerranoThe2016,directem2020}. Some machine learning methods, for instance, embeddings of graphs such as latent space embeddings, Node2vec, and Deepwalk, have found important applications for community detection and link prediction in social networks. Maximilian Nickel and Douwe Kiela present an efficient algorithm to learn the embeddings based on Riemannian optimization~\cite{Nickel2017Poincar}. This method is carried out on the Poincar\'e ball model, as it is well suited for gradient-based optimization (details in Appendix \ref{appendix:effectiveness of hyperbolic embedding}). Here, we embedded the GMNs using a 2-dimensional Poincar\'e disk, with a learning rate of 0.1 and a negative sample size of 30. This setup produced hyperbolic embeddings in which each node $i$---a country in the migration embedding---has radius $r_i$ and angle $\theta_i$. Nodes with small radius hold central positions in the circularly arrayed hierarchy. The hyperbolic distance (depending on angle and radius) between two nodes quantifies their migration relation. Please find the comparative evaluation of hyperbolic embedding and Euclidean embedding in Section \ref{sec:embedding}.

\section{Results and Discussion}

\subsection{Basic statistical characteristics of GMNs}

\begin{figure}[htp]
    \centering
    \subfigure[]{
    \includegraphics[width=0.45\textwidth]{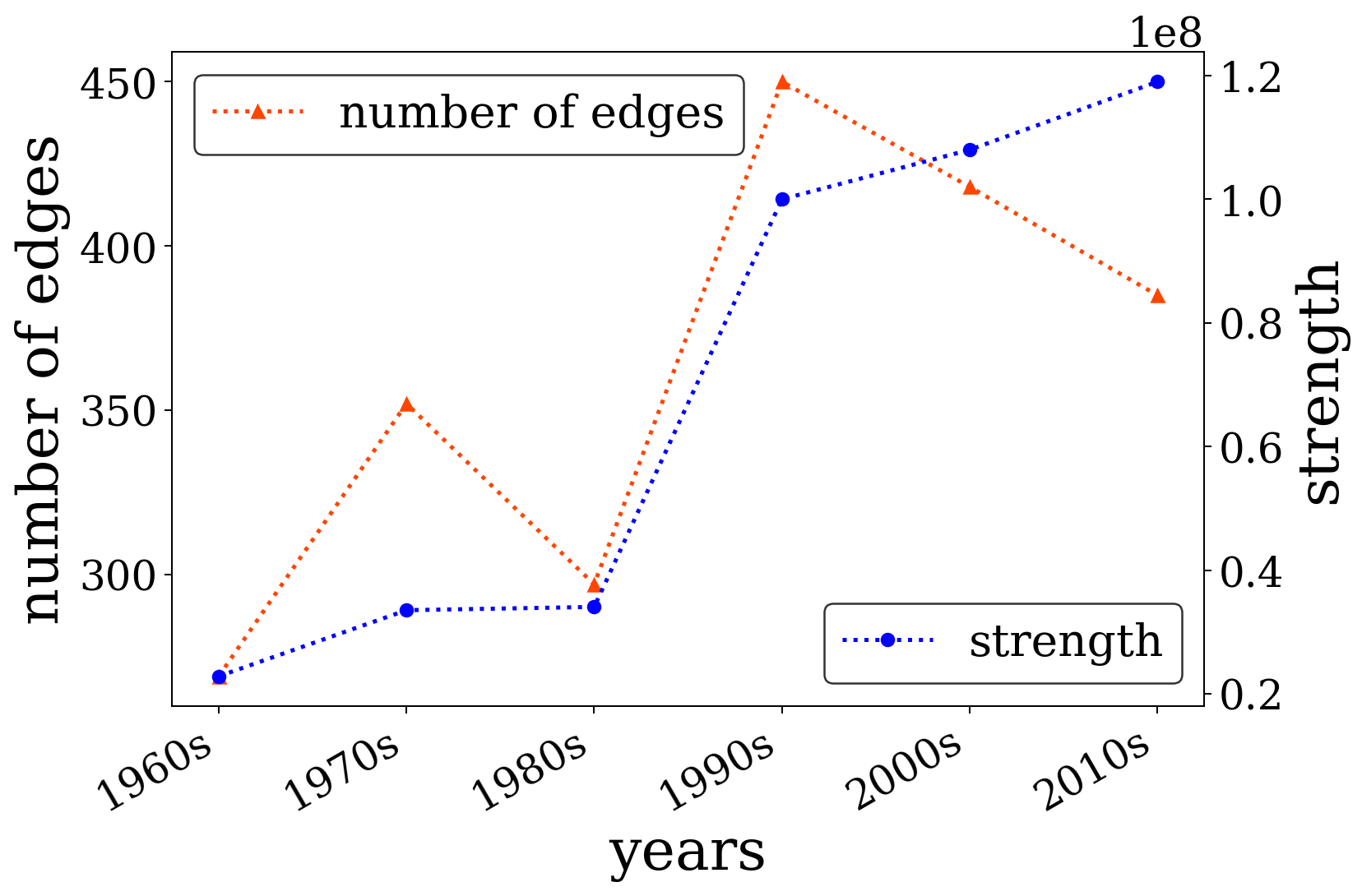}
}
    \subfigure[]{
    \includegraphics[width=0.45\textwidth]{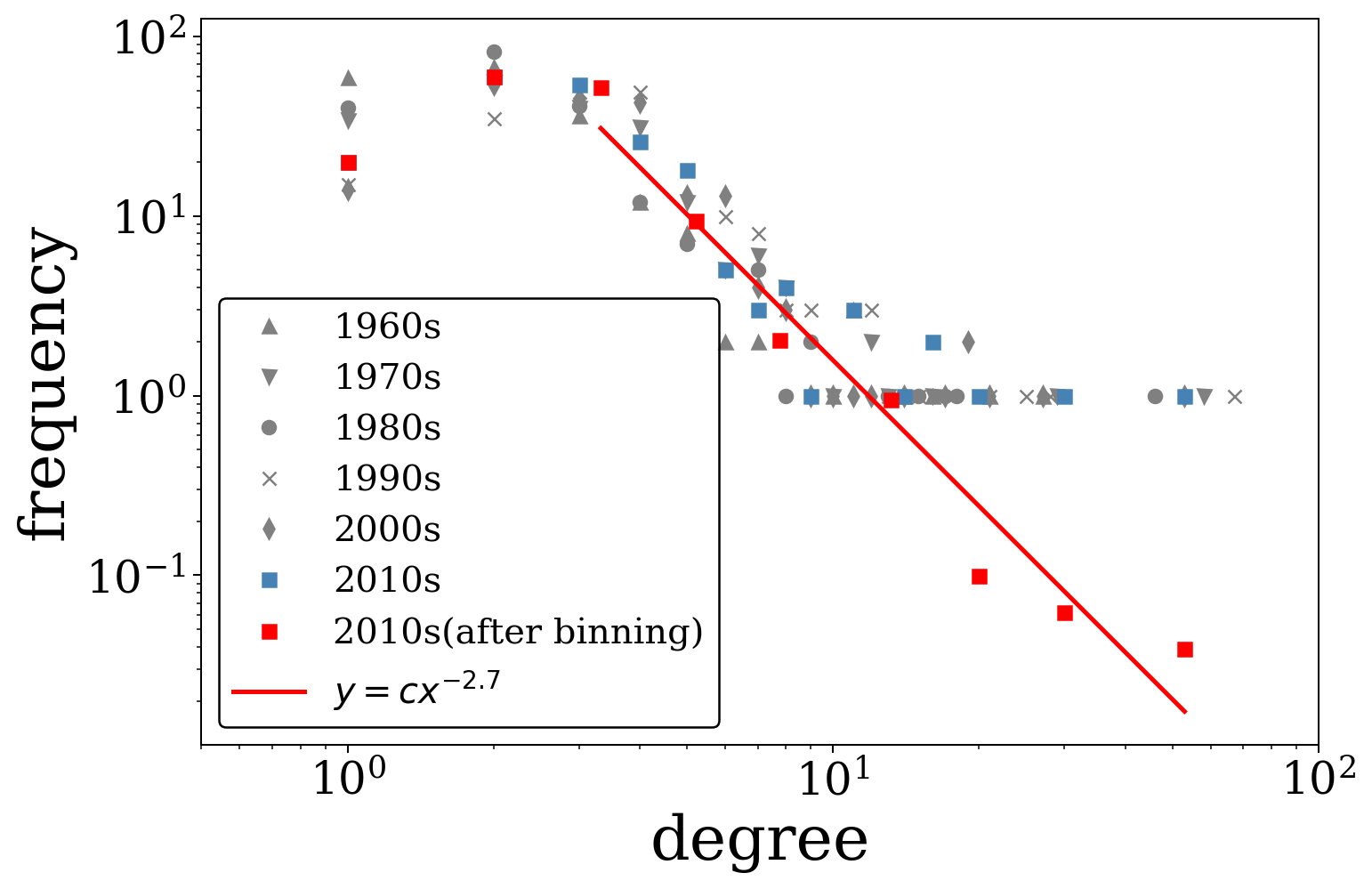}
}
    \subfigure[]{
    \includegraphics[width=0.45\textwidth]{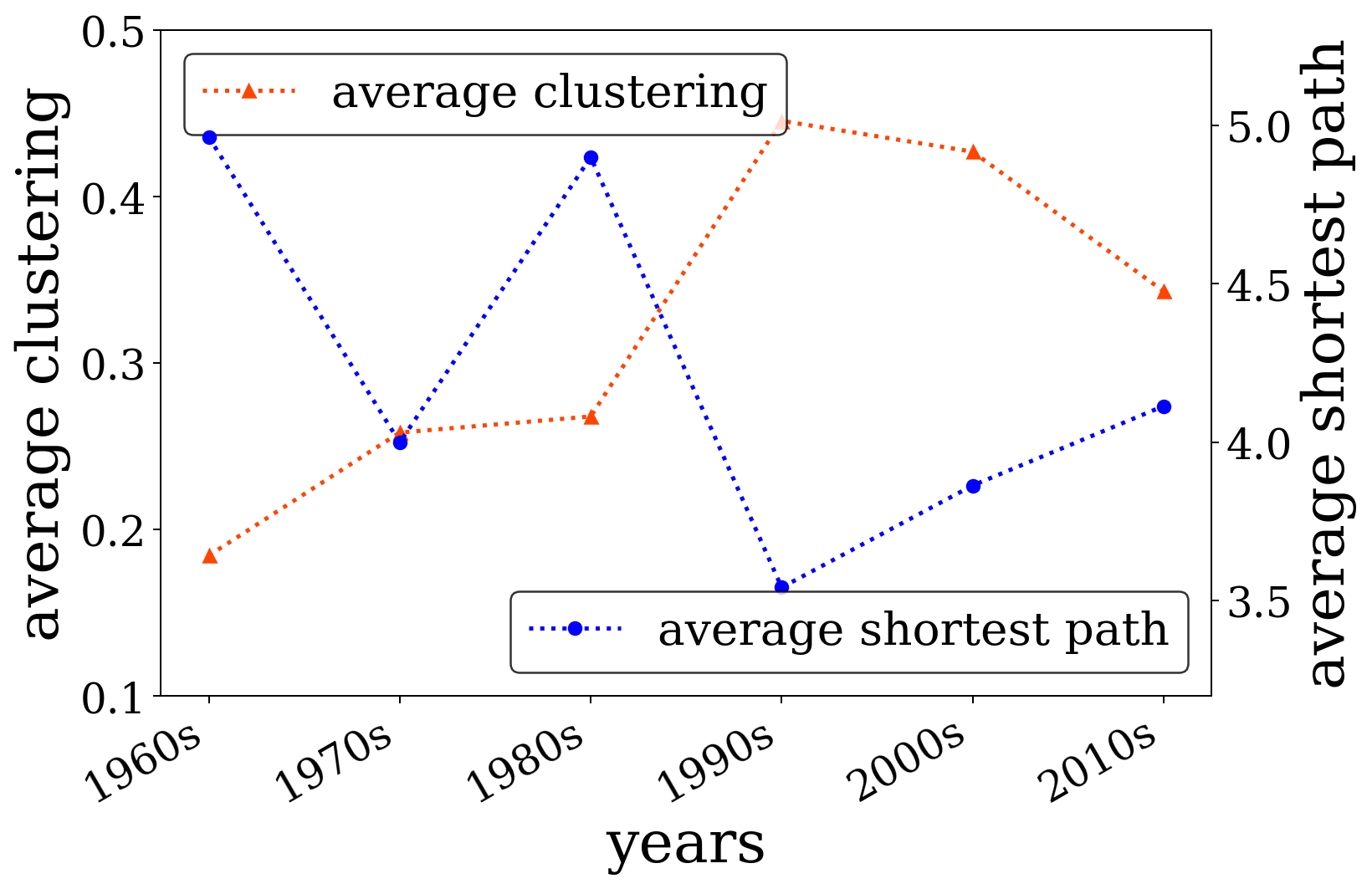}
    }
    \subfigure[]{
    \includegraphics[width=0.45\textwidth]{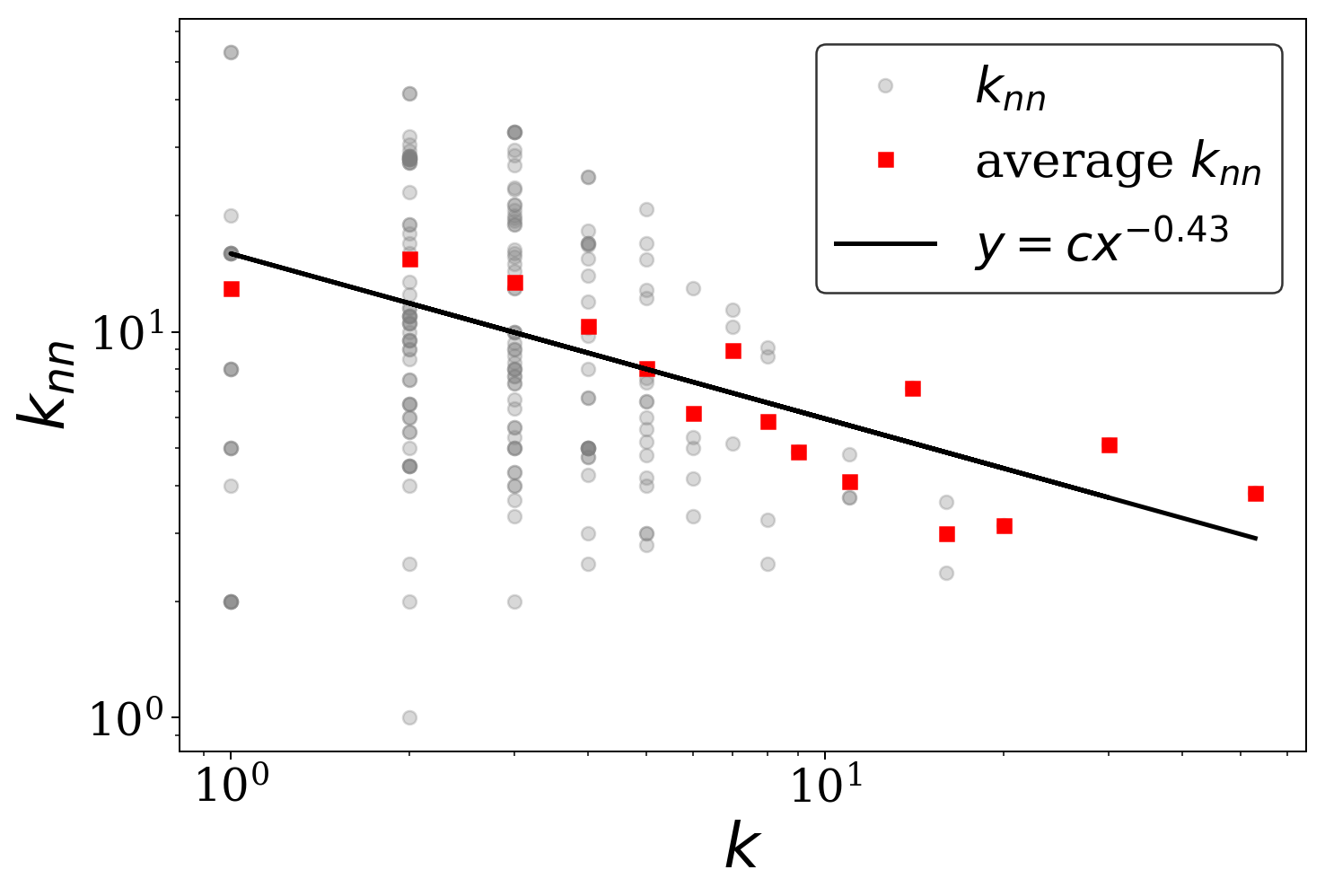}
    }
    \caption{\textbf{(a)} Changes in the strength and numbers of edges of the backbone networks from 1960 to 2015. \textbf{(b)} The degree distribution of the backbone networks and power rate fitting of the population migration network for 2010s. The blue points are the original degree distribution of 2010s, and the red points are the data after binning. \textbf{(c)} Changes in the clustering coefficients and the average shortest path of the backbone networks from 1960 to 2015. \textbf{(d)} Degree correlation of the migration network in 2010s.}
    \label{fig:com_net_ana}
\end{figure}

Fig. \ref{fig:com_net_ana} (a) shows the evolution of the strength and numbers of edges for GMNs from 1960 to 2015. Obviously, the number of edges and the sum of weights exhibit a growth trend over time, which can also be observed in Fig. \ref{fig:backbone}. We believe that this growth implies a more frequent trend of global population migration. You can see the Appendix \ref{appendix:Statistical characteristics of GMN} for the specific statistics of each network. Fig. \ref{fig:com_net_ana} (b) shows the degree distribution of the backbone networks for all periods in gray and blue. We use the backbone network of 2010-2015 as an example to perform power rate fitting for the degree distribution by the nonequidistant bin method, shown in Fig. \ref{fig:com_net_ana} (b) with red dots and line. It was found that the network essentially conformed to the power law distribution. In other words, most countries have a single direction of population migration, while a small number of countries have population exchanges with many countries; this means that population migration exhibits local concentration.

The clustering coefficient measures the degree of connectivity of a network; the average shortest path reflects the difficulty of one node connecting to another node in the network. The clustering coefficients and the average shortest path (in the maximum connected subnet) of GMNs are shown in Fig. \ref{fig:com_net_ana} (c). The clustering coefficient exhibits an upward trend, while the shortest path follows a downward trend, exhibiting the enhancement of small world attributes. In our opinion, the global migration relations have become closer in recent years, which also indicates the increase in network clustering during 1960-2015.

The degree correlation in complex networks reflects the connection preference of nodes in the network, as defined below:

\begin{equation}
    k_{nn}(k)\equiv \sum_{k'}k'P(k'|k) 
\end{equation}
with scaling hypothesis $k_{nn}(k)\propto k^{\mu}$. $k_{nn}$ indicates the average degree of the first neighbors of nodes with degree $k$. If $\mu >0$, then this is an assortative network; similarly, $\mu =0$ indicates a neutral network, while $\mu <0$ indicates a disassortative network. We calculate the degree correlation of each network by degree correlation function, and their $\mu$ values are all negative during 1960-2015. Fig. \ref{fig:com_net_ana} (d) shows the degree correlation of the 2010s. The GMNs in other years exhibit similar imagines. This method shows that all of the networks have the characteristics of negative matching, indicating that nodes with lower degrees are more likely to be connected with nodes with higher degrees. The reason may be that the migration of most small countries shows preferential movement to several large countries that have survival advantages or economic advantages. In fact, the disassortative characteristic of the international migration network has been verified in some existing studies~\cite{porat2016global,International-MigrationNetworke}. In addition to the migration networks, the international oil trading network shows a disassortative feature in which countries with fewer trading partners tend to develop oil trading relations with countries with more trading partners~\cite{oiltrade}. 

\subsection{Hyperbolic characteristics of GMNs}
\label{sec:embedding}

Some network structures could actually be better described by hyperbolic space~\cite{Michele2015Hyperbolicity,SerranoThe2016}. To further compare the advantages of hyperbolic embedding, we try to separately embed GMNs into Euclidean and hyperbolic planes. We consider the least squares error function used in \cite{errorfun2017}. After unified measurement, the errors of two spaces are revealed in Fig. \ref{fig:error&score} with dotted lines. In addition, we also compute the embedding $score$ (details in Appendix \ref{appendix:effectiveness of hyperbolic embedding}) in both two spaces (considering the error function performance, we only embed the network in Euclidean space by nonclassical MDS). The result is shown in Fig. \ref{fig:error&score} with solid lines. This figure shows that all errors in hyperbolic embedding are lower than those in Euclidean embedding; according to the $Score$, hyperbolic embedding also offers a more professional performance in the expression of data size relations. This result indicates that the GMNs exhibit a significant hyperbolic characteristic during 1960-2015.

\begin{figure}
    \centering
    \includegraphics[width=0.6\textwidth]{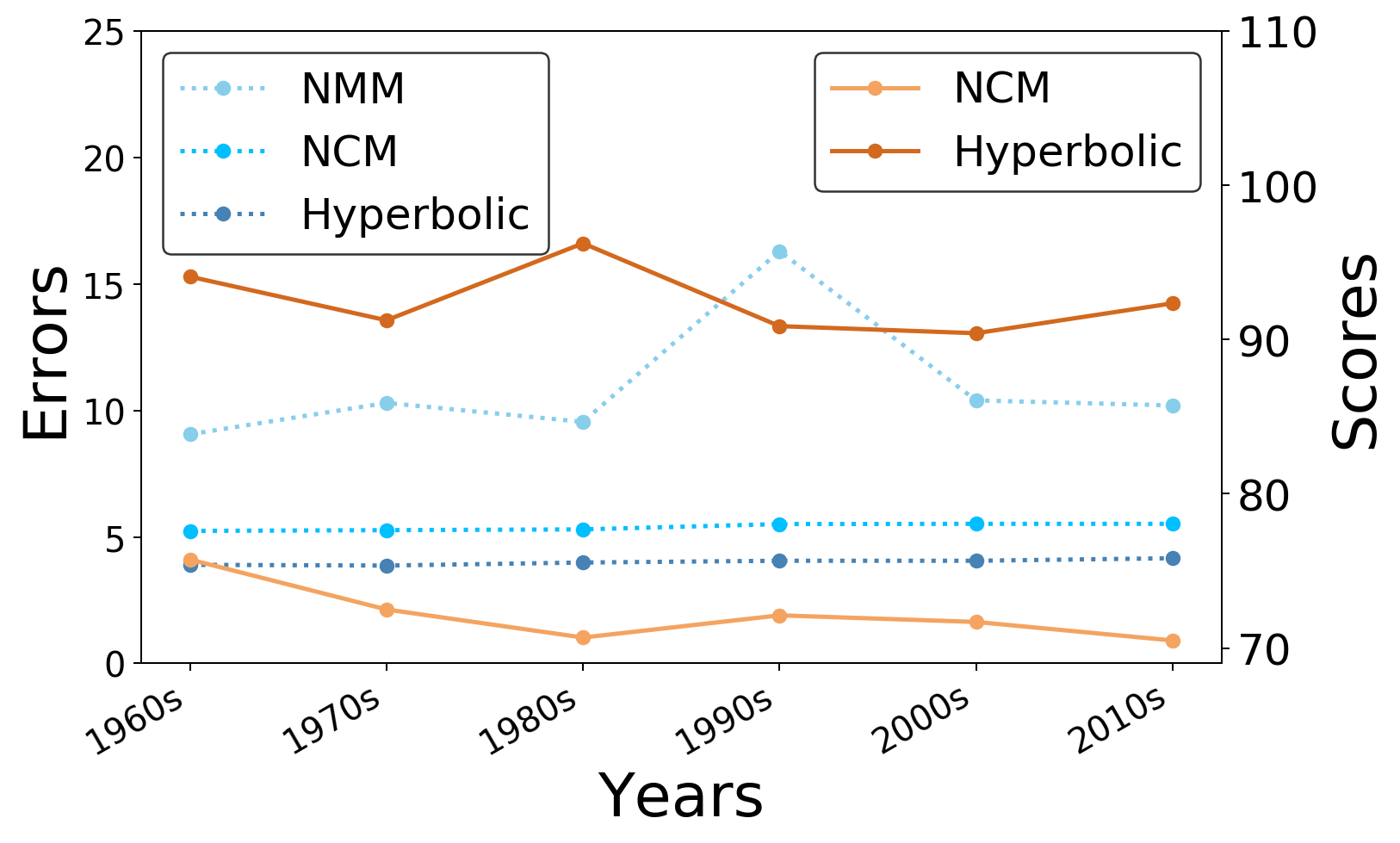}
    \caption{The error functions and embedding scores of GMNs in Euclidean and hyperbolic planes for 1960-2015.}
    \label{fig:error&score}
\end{figure}

\begin{figure}[htp]
    \centering
    \subfigure[1960-1969]{
     \includegraphics[width=5.5cm]{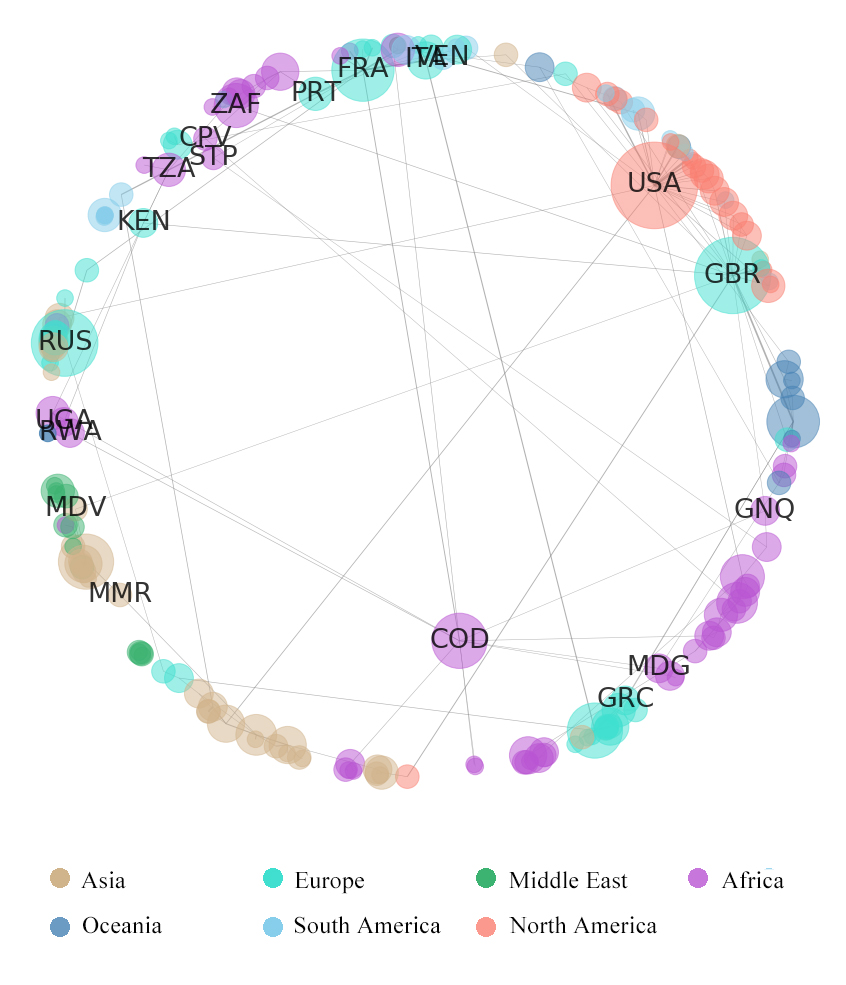}
     }
   \subfigure[2010-2015]{
     \includegraphics[width=5.5cm]{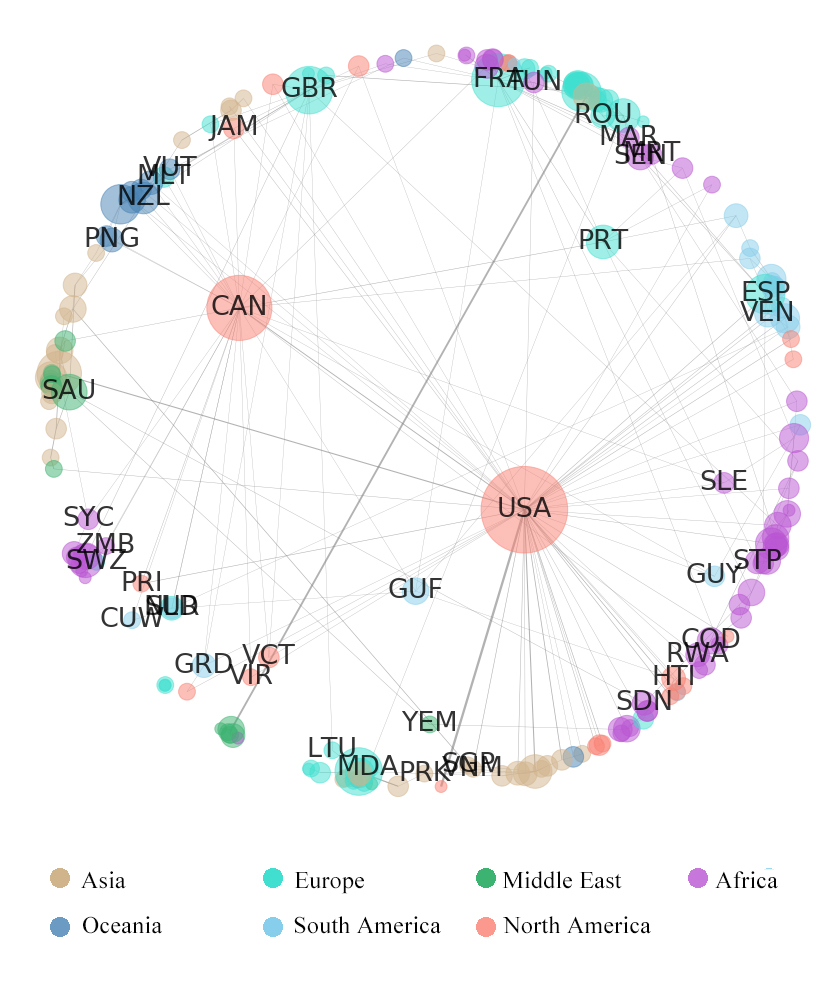}
     }
    \caption{Hyperbolic embedding map of global migration network.}
    \label{fig:HyEm2015}
\end{figure}

The Poincar\'e disk embedding results of GMNs in 1960s and 2010s are visualized in the Fig. \ref{fig:HyEm2015}. Each node represents a country/region. The size of the node expresses its degree in GMN, the hyperbolic distances represent their relations in the global migration network. For clarity, only the names of nonperipheral countries whose distance less than 0.97 from the origin of the coordinates are shown. The color of the node indicates the geographical location of the country/region. The figure indicates that GMNs presents an obvious ``core-periphery" structure. In the 1960s, the network is sparse, while in the 2010s, 200 countries have closer and more complex migration relations, which also conforms to the common law of global integration. The population migration in the 1960s primarily occurred within the regions or communities, and there are fundamentally fewer typical countries in the center of the Poincar\'e disk (Fig. \ref{fig:HyEm2015} (a)). The Democratic Republic of the Congo (COD) is the second largest country in Africa. Once a Belgian colony, it became independent in 1960 and became a link in the global migration network between France (FRA) in Europe and African countries such as Sudan (SUD) and Madagascar (MDG).

Furthermore, for the period of 2010-2015, the network structure is more complex, and the hierarchy is more obvious. Here, the United States and Canada (with large degrees of $k=53$ and $k=30$, respectively), which have closer population migration relations with other countries in various regions of the world, are more centrally located in the disk. France and the United Kingdom, which mainly connect local communities such as European and some African countries, have been slightly more marginal since the 2010s. Although the degrees of Portugal and Yemen are not large ($k=8$ and $k=2$, respectively), their locations indicate their contributions on connecting the migration of several continents. 

Additionally, Fig. \ref{fig:HyEm2015} also shows that the hyperbolic distance between countries/regions is not entirely determined by the geographical location. It represents the correlation of embedding distances and geographic distances, which is significant during 1960-2015 (with sig.$\approx 0$). This means that hyperbolic distance $d_h$ is positively correlated with geographical distance $d_g$ but encodes more than purely geographical information. 

\begin{figure}[htp]
    \centering
    \subfigure[Definition of neighborhood]{
     \includegraphics[width=0.52\textwidth]{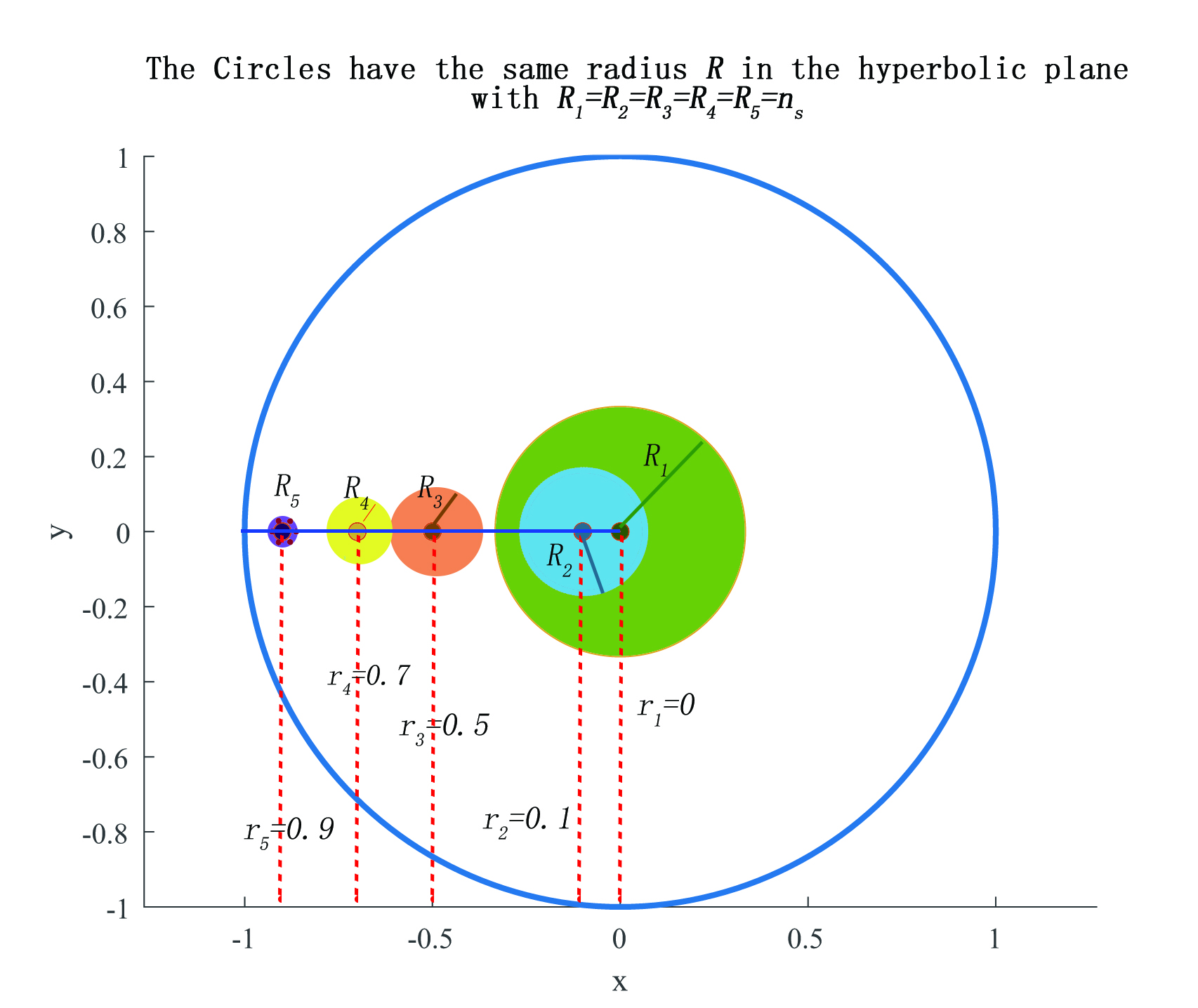}
     }
   \subfigure[Angular density]{
     \includegraphics[width=0.43\textwidth]{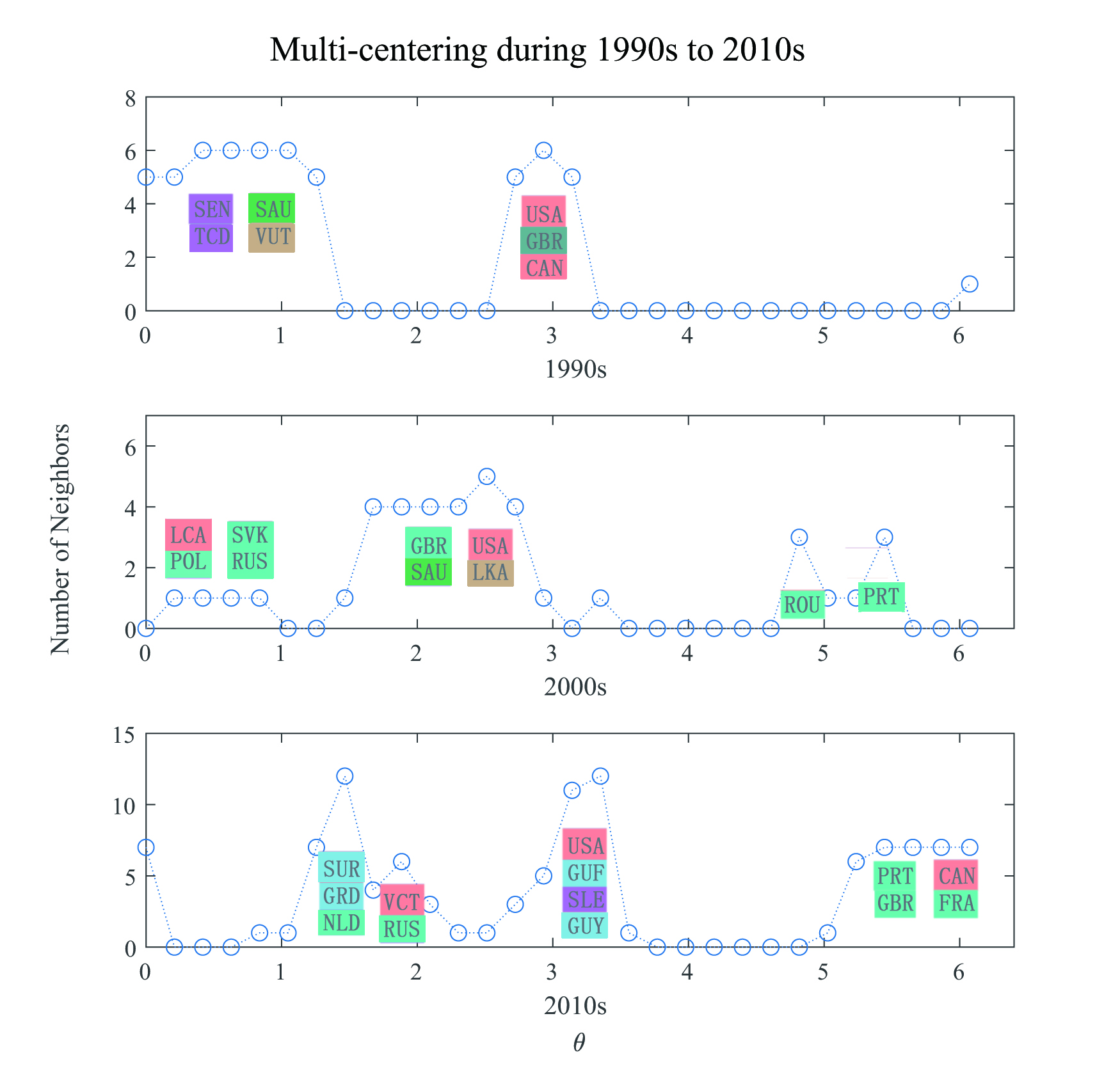}
     }
    \caption{Angular density in hyperbolic plane which could describes the aggregation of countries and the multi-centering characteristics of global migration networks.}
    \label{fig:angular density}
\end{figure}

In order to describe the distribution characteristics of the countries on the hyperbolic plane more clearly, we define the angular density $f(\theta)$. First of all, we get some points on the hyperbolic plane (the points with polar coordinates of $[\pi,0],[\pi,0.1],[\pi,0.5],[\pi,0.7],[\pi,0.9]$ in Fig. \ref{fig:angular density} (a)); then we draw neighbourhood of the points with the same radius $n_s$. As these points move away from the core ($r=0$), their neighborhood looks smaller, and the center of the circle tends to shift toward the core, but on the hyperbolic plane, these circles have the same area. So the green circle with the center $[\pi,0]$, blue with $[\pi,0.1]$, orange with $[\pi,0.5]$, yellow with $[\pi,0.7]$ and purple with $[\pi,0.9]$, they have the same area. And we define the density of points as the number of countries loaded in their neighborhood.

Then we sum the density for each $\theta\in[0,2\pi)$ and get the angular density distribution $f(\theta)$. Angular density can more intuitively show the distribution characteristics of countries in hyperbolic space. Fig. \ref{fig:angular density} (b) shows the angular density for 1990s-2000s. Blue dotted line indicates the angular density for each $\theta$, and the countries are the representatives with the relatively centering positions and they belongs to the corresponding peaks. The color of the country indicates its region.

Angular density can help us see the aggregation of the global migration networks more intuitively: 1. In 2010s, the aggregation has increased obviously, which is reflected by the increasing angular density, and more countries have gathered together. 2. Most of the countries in the center are located in North America or Europe. It is mainly because the countries in North America and Europe have the shorter migration distance from other countries, so they are naturally easier to be located in the center of hyperbolic plane. 3. Besides, in 1990s, African countries were more aggregated, while in 2010s, Latin American countries were more aggregated. 4. It shows a trend of ``multi-centering'' of the global migration networks since 1990s.

\subsection{Communities characteristics of GMNs}
\label{sec:communities}

\begin{figure}[htp]
    \centering
    \subfigure[]{
    \includegraphics[width=0.45\textwidth]{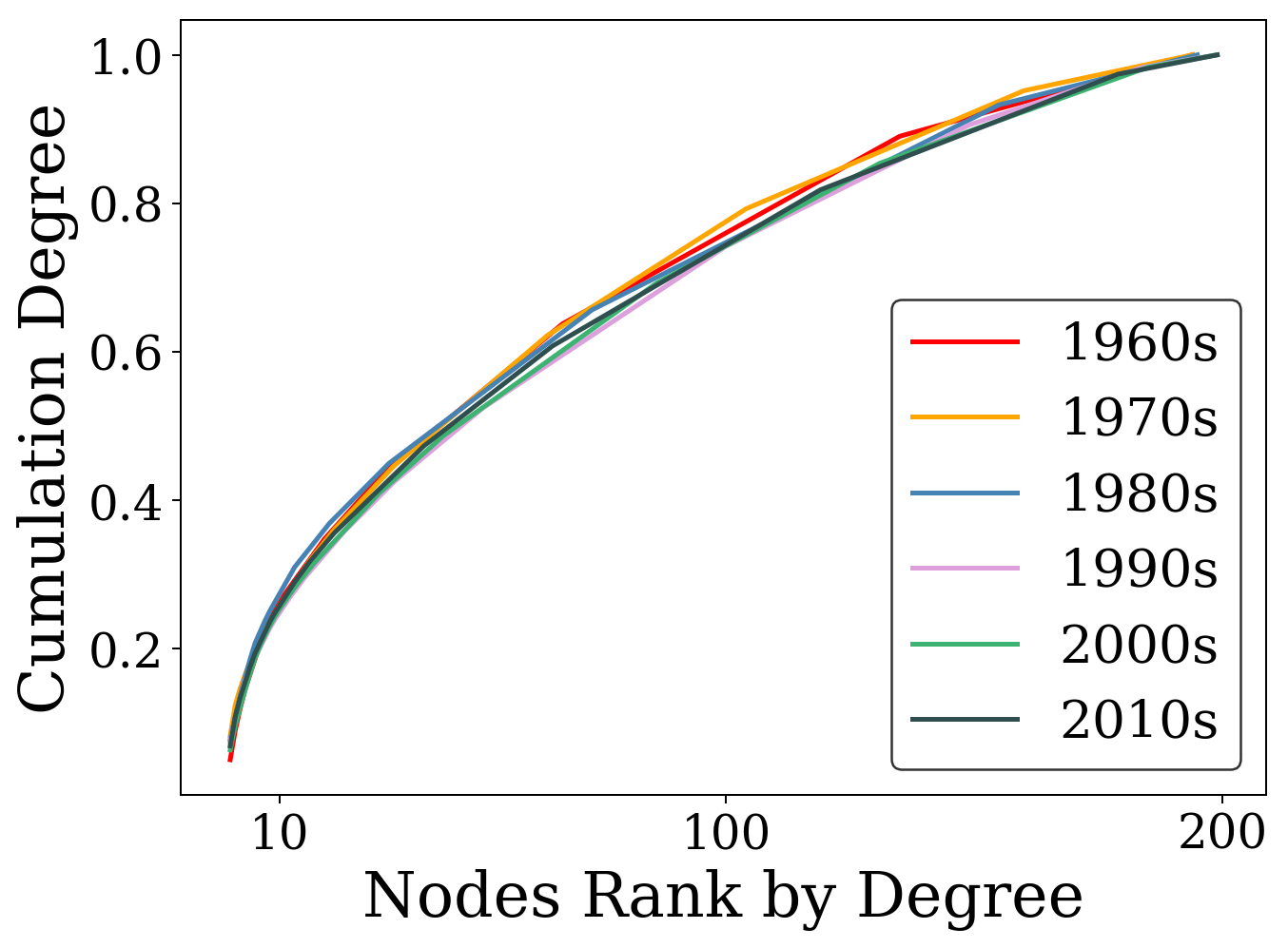}
}
    \subfigure[]{
    \includegraphics[width=0.45\textwidth]{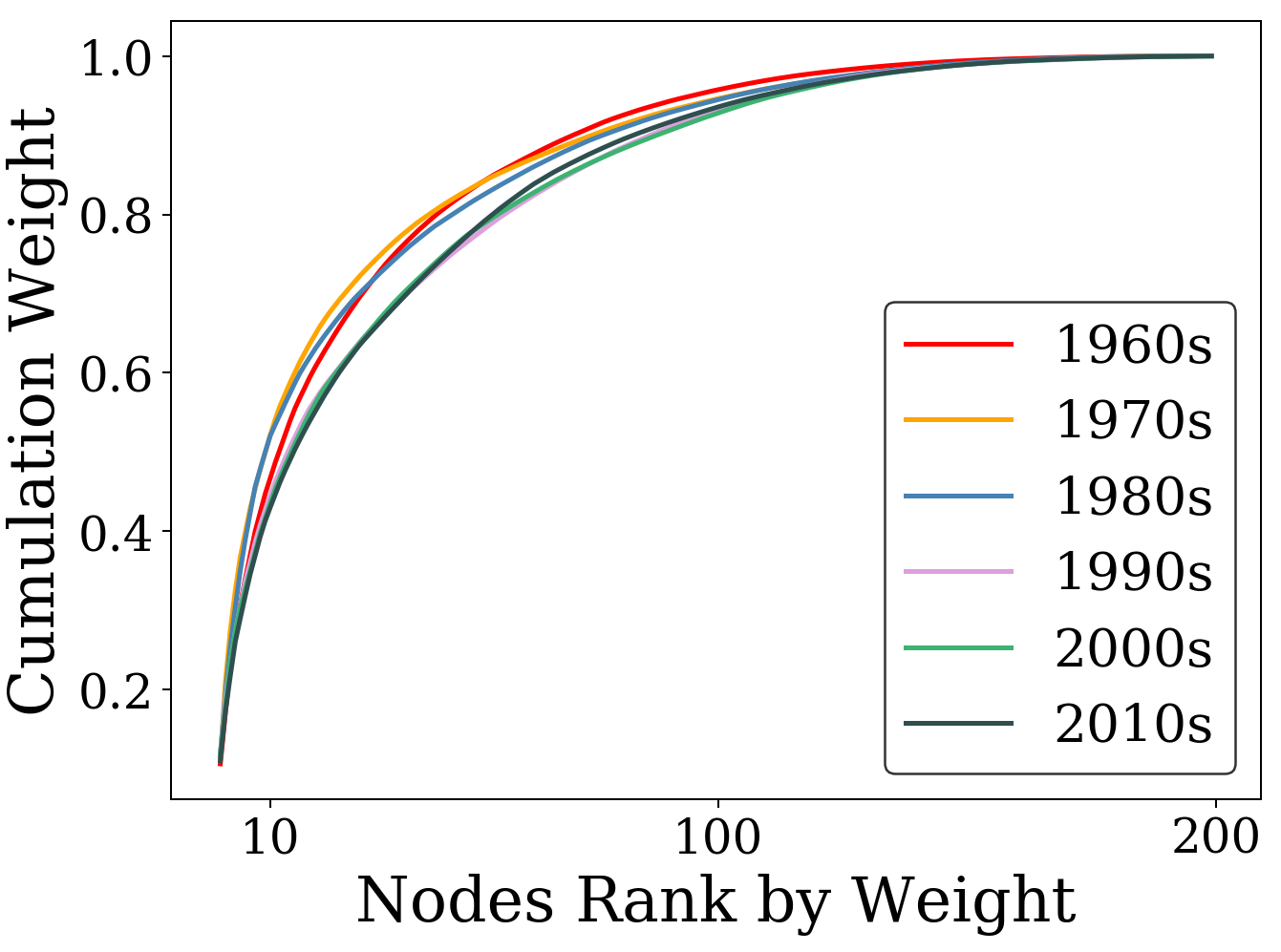}
}
    \subfigure[]{
    \includegraphics[width=0.45\textwidth]{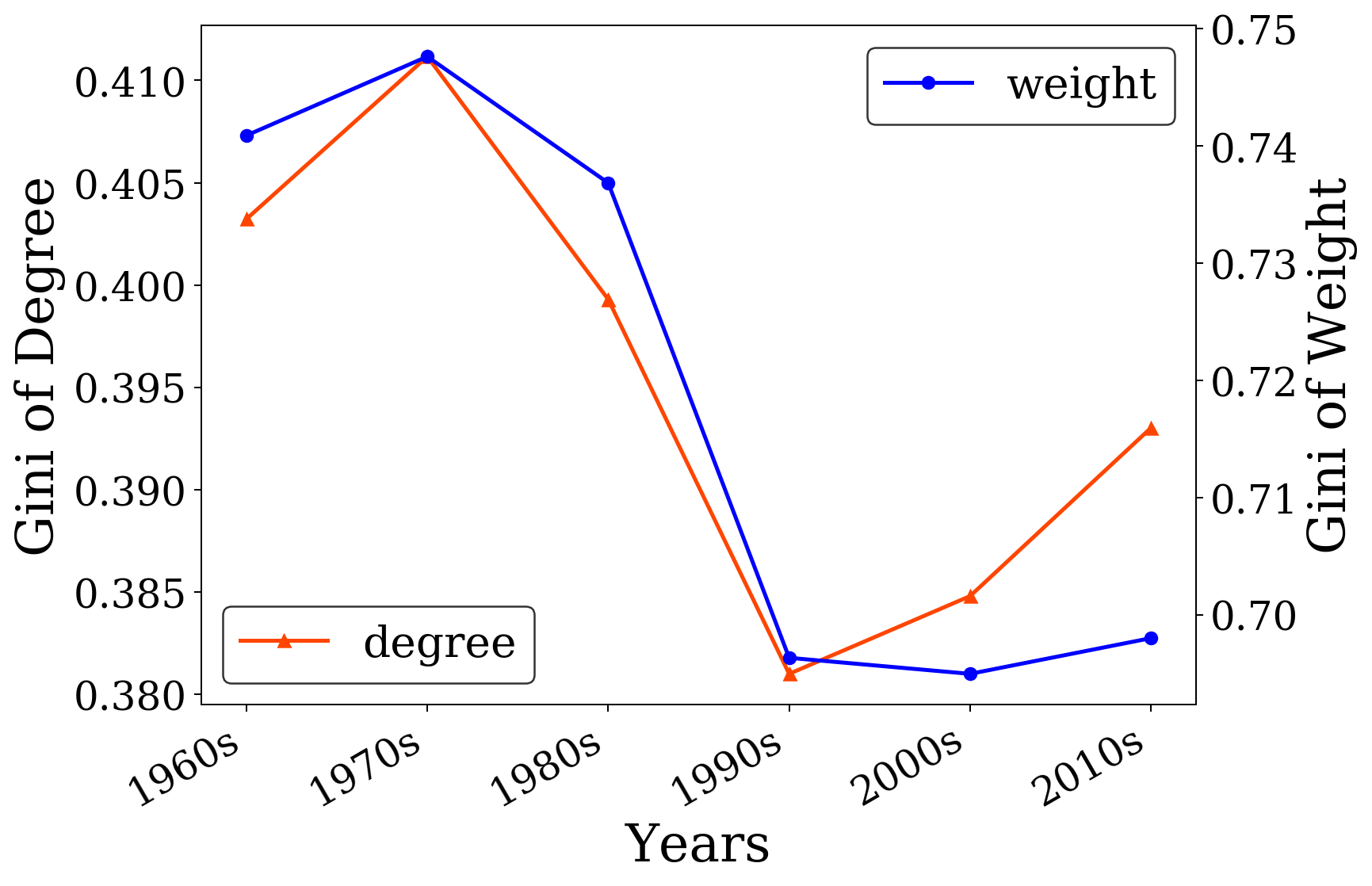}
    }
    \subfigure[]{
    \includegraphics[width=0.45\textwidth]{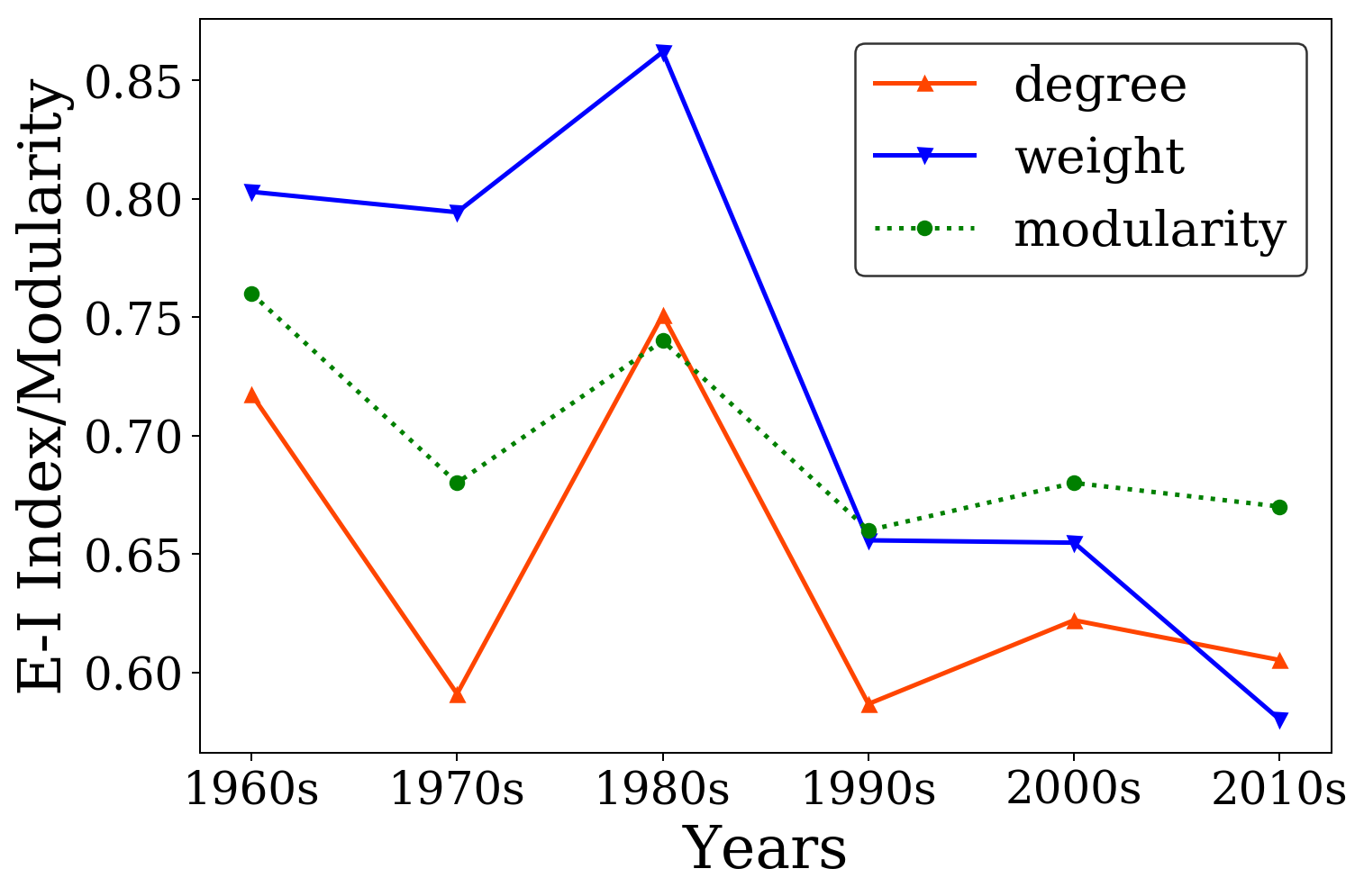}
    }
    \caption{\textbf{(a)} The cumulative distribution of degrees for nodes from 1960 to 2015. \textbf{(b)} The cumulative distribution of weights for nodes from 1960 to 2015. \textbf{(c)} Changes in the Gini index for the degrees and weights from 1960 to 2015. \textbf{(d)} Changes in the E-I index from 1960 to 2015.}
    \label{fig:EI}
\end{figure}

There are many ways to explore the communities of a complex network, such as the GN algorithm~\cite{GN2004} based on network topology and Potts model~\cite{PottsModel2004} based on network dynamics. In this paper, the Louvain algorithm~\cite{louvain2008} based on modularity, which is rapid and exhibits an obvious clustering effect, is adopted. The algorithm divides each round of calculation into two steps: in the first step, the algorithm scans all nodes, traverses all neighbors of the node, and measures the modularity benefit of adding the node to the community of its neighbor; it then selects the corresponding neighbor node with the highest modularity gain and joins its community. This process is repeated until the results are stable. During 1960-2015, the modularity value $Q$ is within 0.66-0.76, which proves the validity of clustering (green line in Fig. \ref{fig:EI}(d)).

\begin{figure}[htp]
    \centering
    \subfigure[1960-1969]{
    \includegraphics[width=0.45\textwidth]{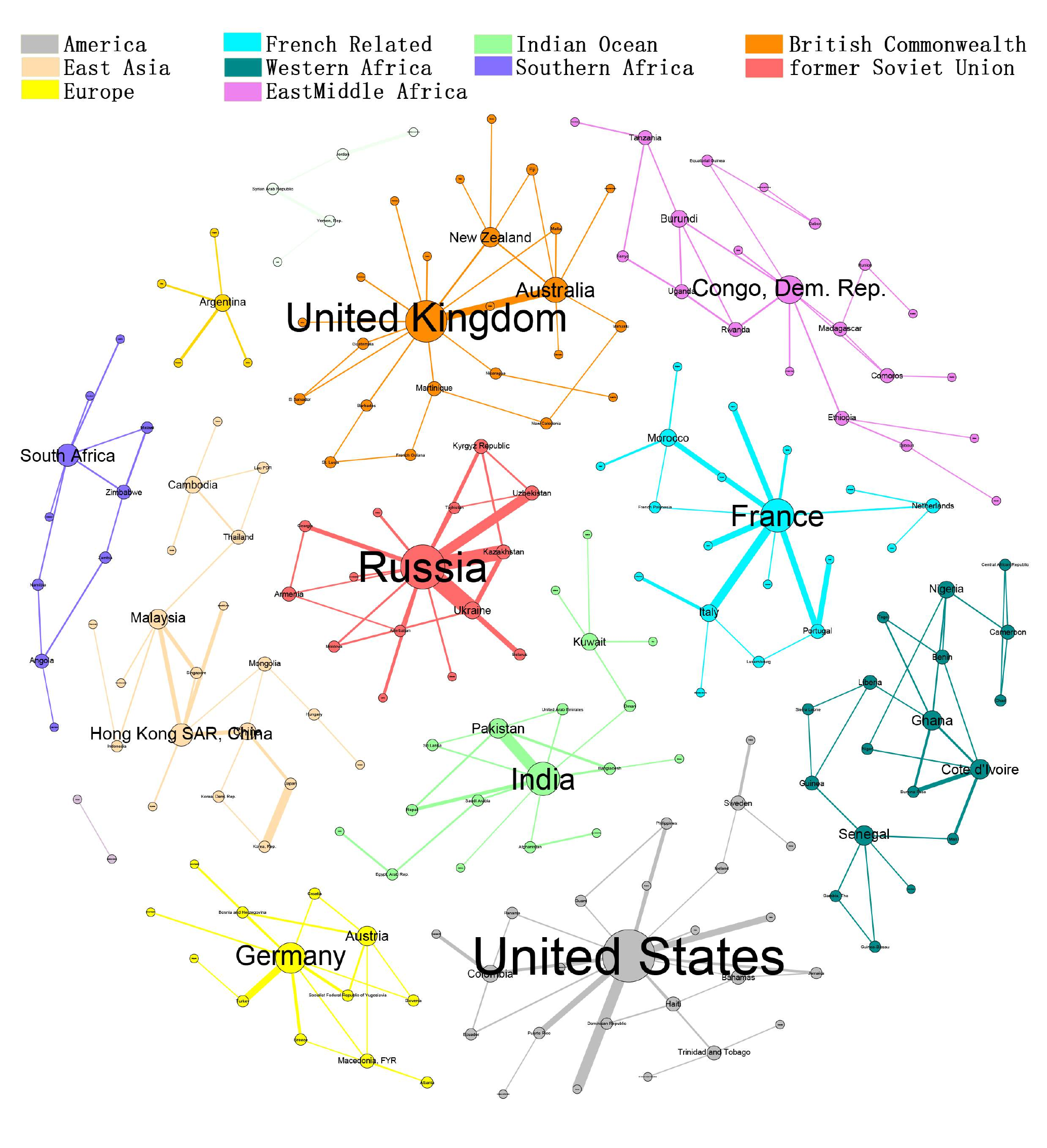}
}
    \subfigure[1970-1979]{
    \includegraphics[width=0.45\textwidth]{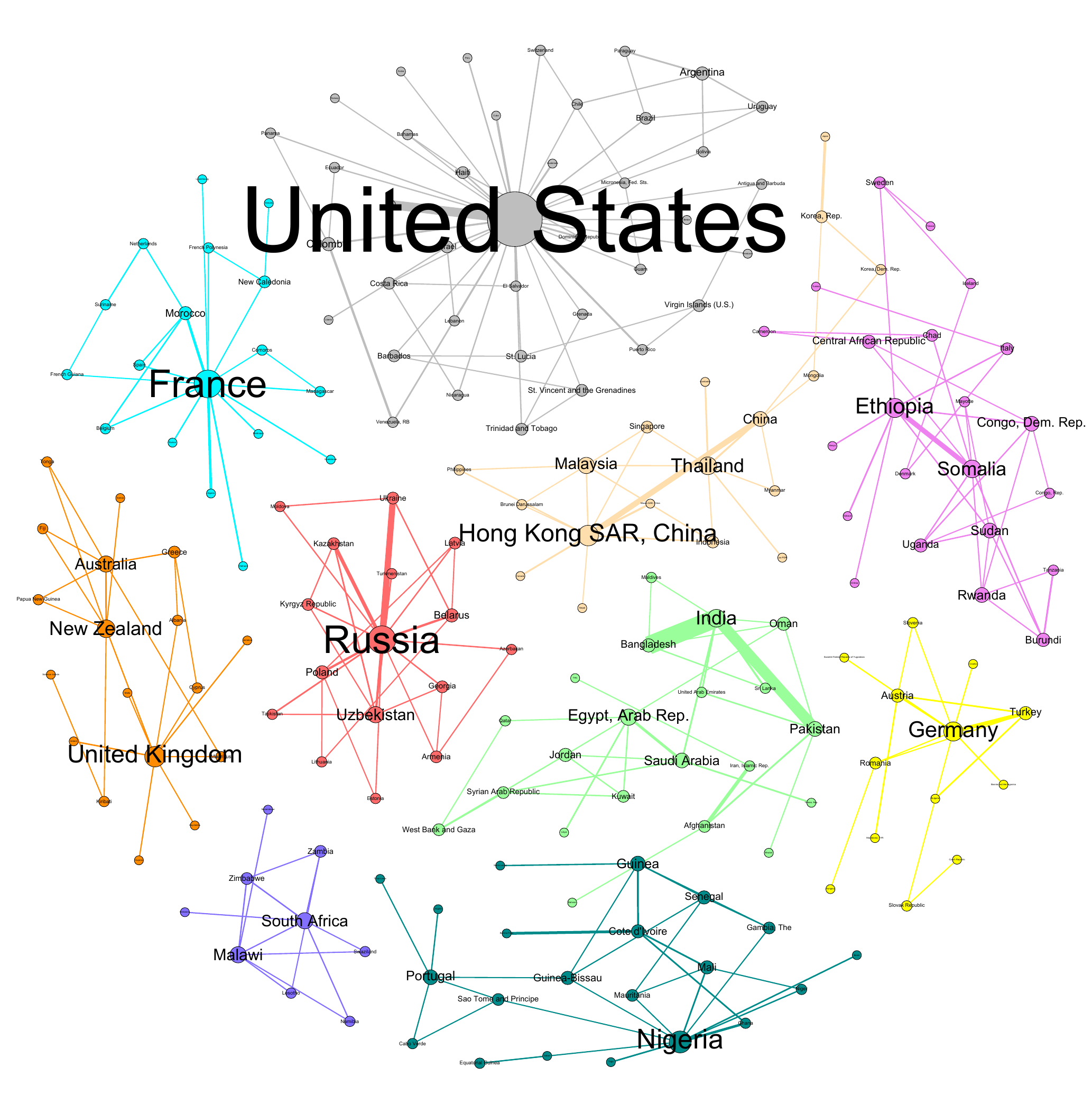}
}
    \subfigure[1980-1989]{
    \includegraphics[width=0.45\textwidth]{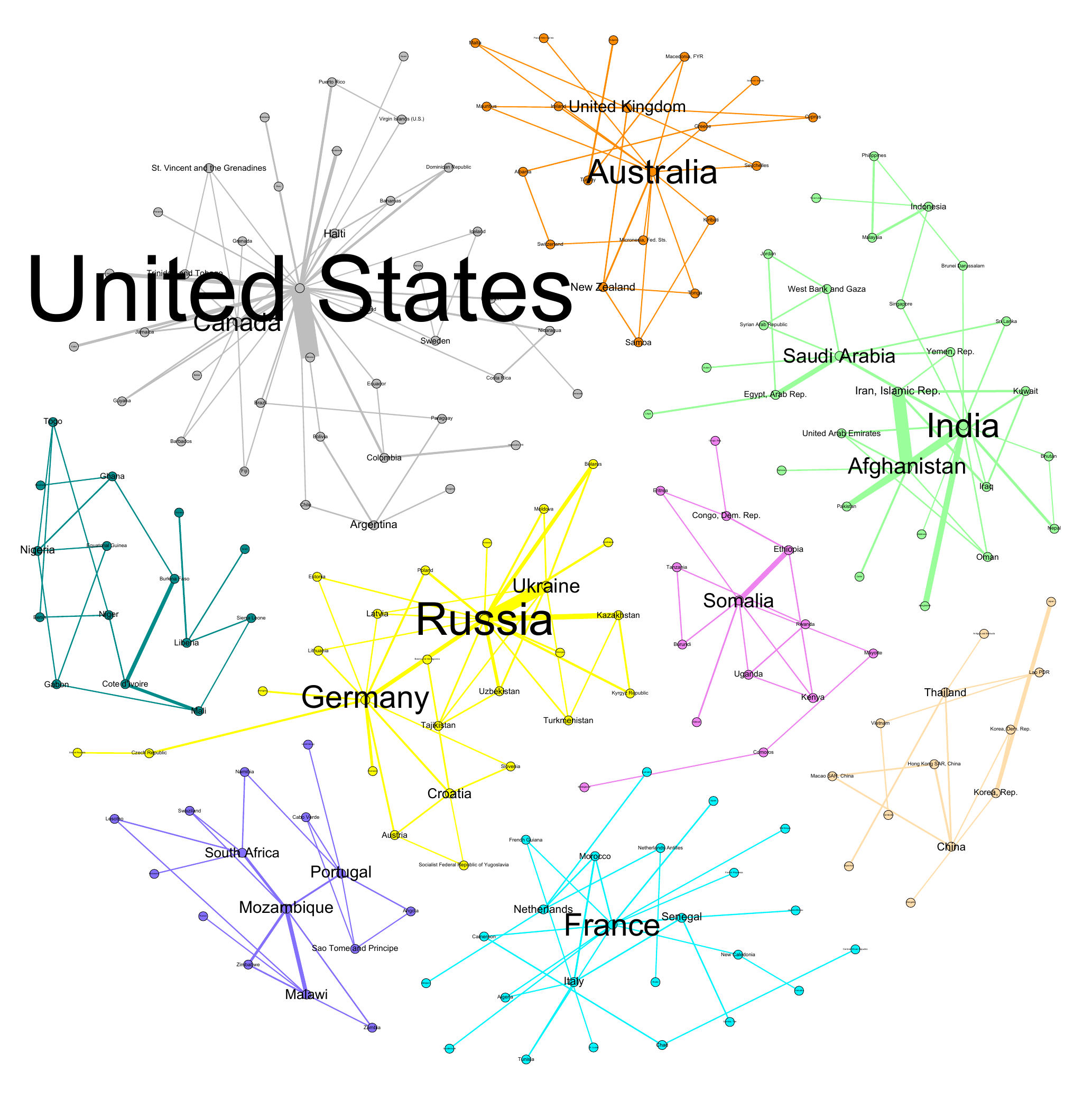}
}
    \subfigure[1990-1999]{
    \includegraphics[width=0.45\textwidth]{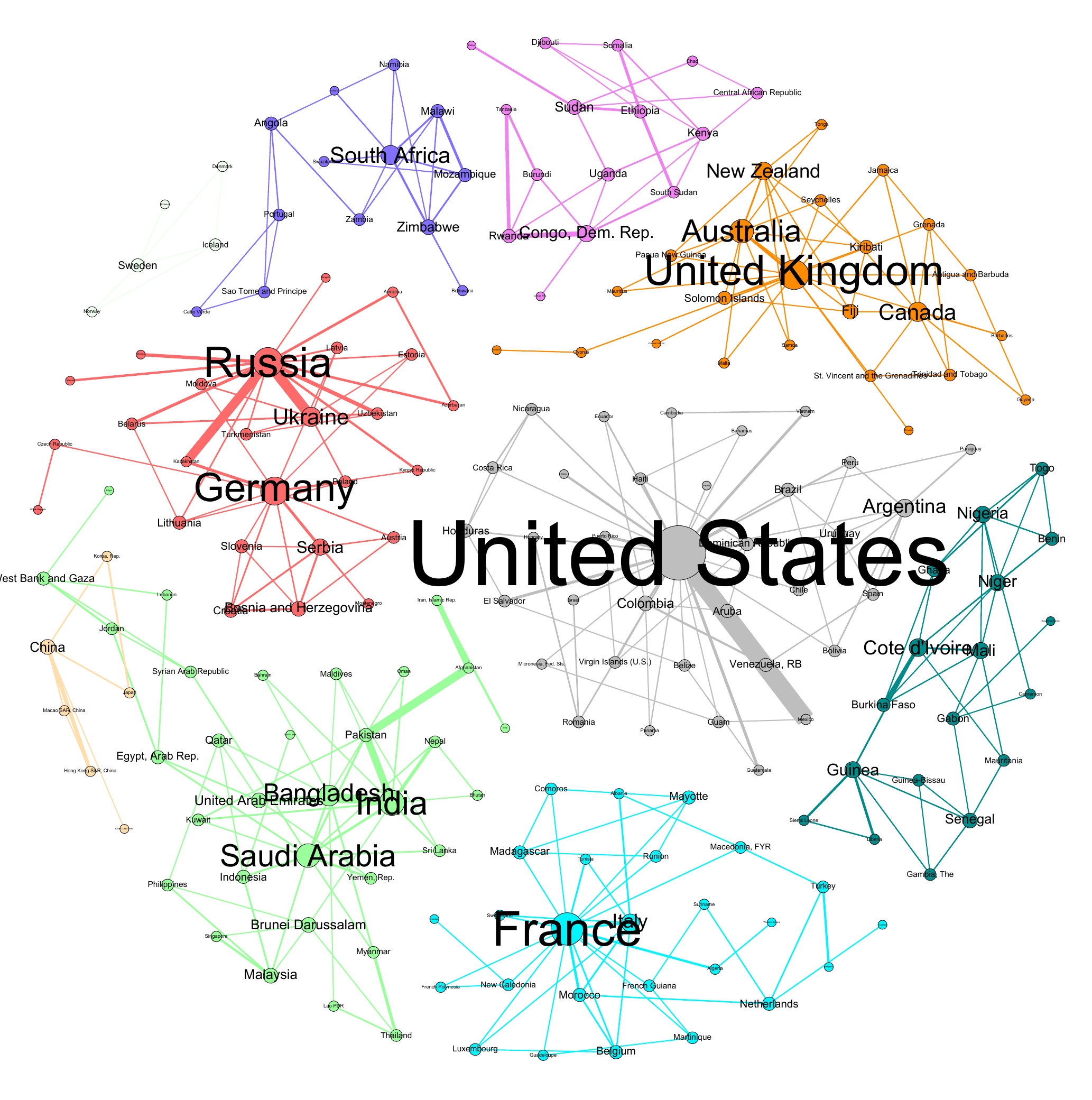}
}
    \subfigure[2000-2009]{
    \includegraphics[width=0.45\textwidth]{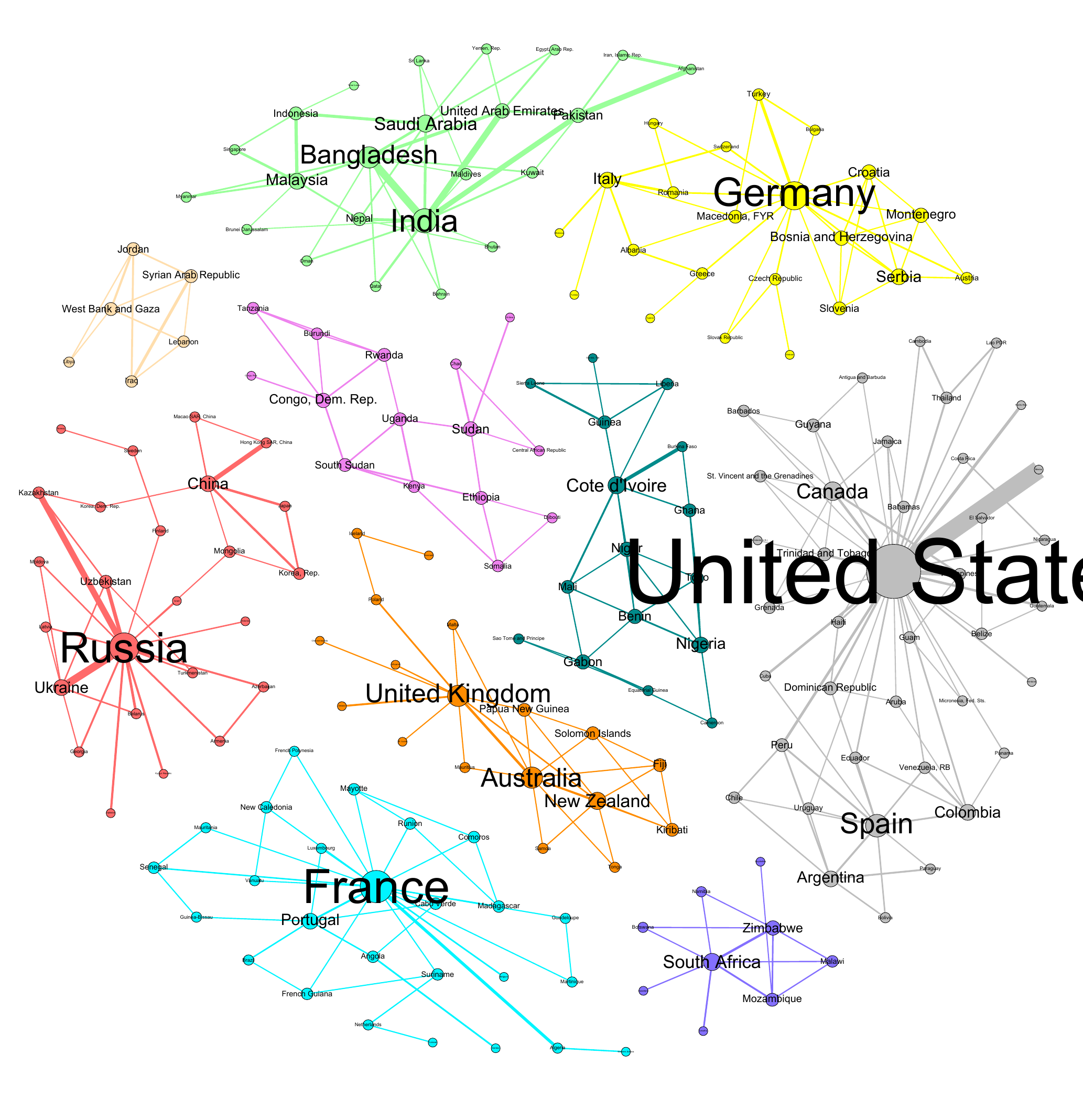}
}
    \subfigure[2010-2015]{
    \includegraphics[width=0.45\textwidth]{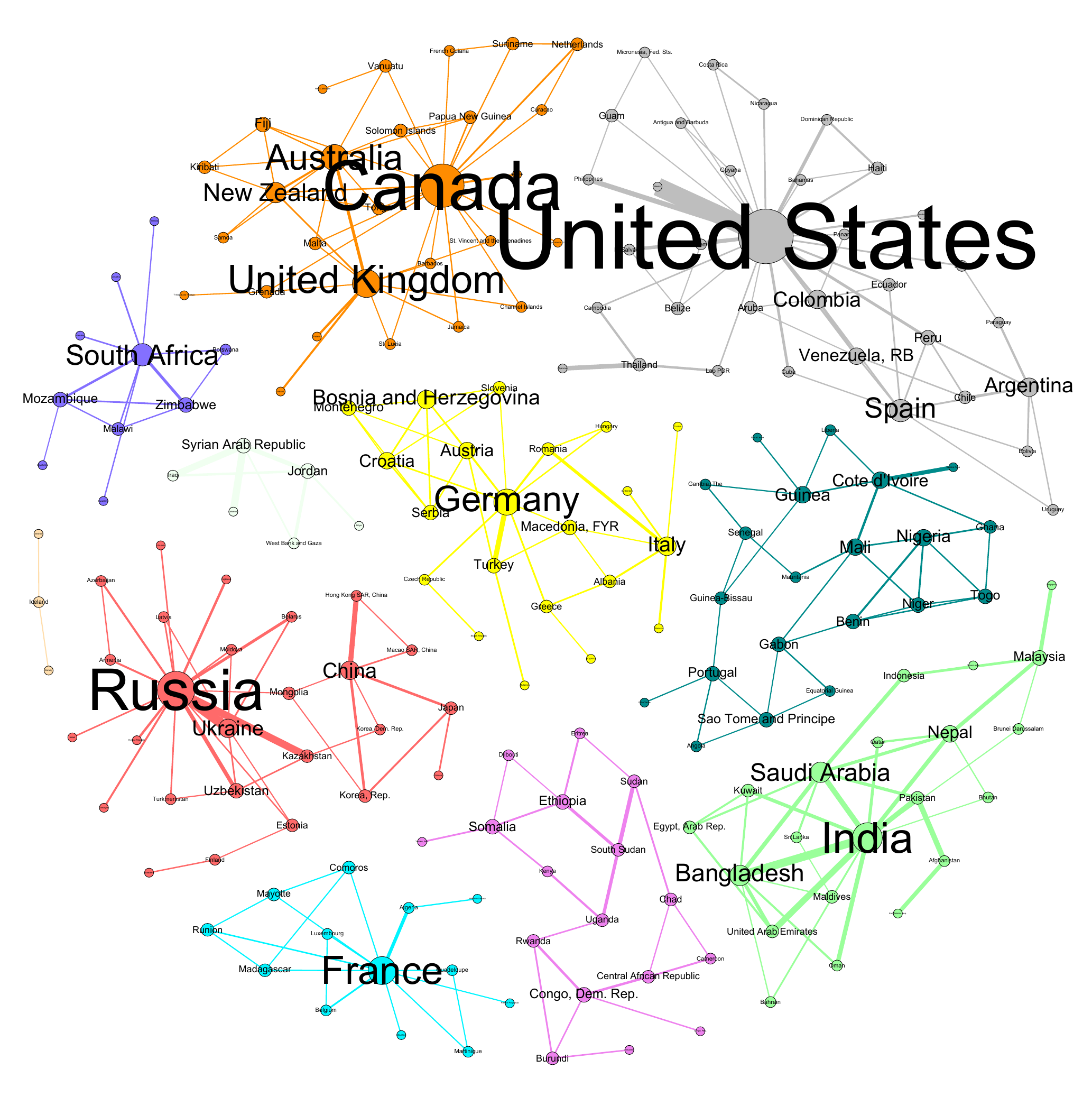}
}
    \caption{Communities and structural evolution of GMNs from 1960 to 2015.}
    \label{fig:cluster}
\end{figure}

Fig. \ref{fig:cluster} shows the communities of global migration networks during 1960-2015. Obviously, the result of clustering is relatively stable over the most recent fifty years. According to the composition of members, we define ten typical communities. 1. America: including the countries in North America, Central America, South America and the Caribbean; 2. French related: including France and other neighboring European countries, as well as French territories and former colonies around the world; 3. British Commonwealth: including some original and current Commonwealth countries, including Australia, New Zealand, Canada, etc.; 4. Indian Ocean: centered on India, including South Asia, North Africa, Southeast Asia and other countries close to the Indian Ocean; 5-7. most
sub-Saharan African countries are divided into three communities: East-Middle, Western and Southern Africa; 8. Europe: some countries in western, central and eastern Europe; 9. former Soviet Union: including Russia and some former Soviet countries; and 10. East Asia: mainly East Asian countries, including some Southeast Asian countries in early times, and Russia and some former Soviet countries in recent years.

After grouping, we could analyze the ``globalization'' or ``polarization'' trends based on comparing the global and local connectivity in migration communities. The cumulative distribution of degrees and weights (Fig. \ref{fig:EI} (a) and (b)) shows that, on the whole, they both tend to be relatively flat. This means that many top countries are reducing their proportions of edges and flows, while other countries with low flows are experiencing relatively rapid development. The Gini coefficients of degree and weight both exhibit an overall downward trend, and the entire network becomes more balanced over time (Fig. \ref{fig:EI} (c)).

In recent years, some scholars have proposed the ``globalization of migration'' hypothesis and emphasized both the progressively increasing number of countries involved in global migration and the diversification of origins and destinations~\cite{Galeano2011Castles,Valentin2016}. Some other scholars offered an alternative understanding, suggesting that in recent years, globalization did not contribute to an overall increase in mobility possibilities but instead widened the gap between rich and poor countries~\cite{Zolberg1989The,Hirst2009Globalization,Hallwarddriemeier2017Trouble}, leading to polarization of the global migration networks.

Here we use the external-internal index (E-I index) to measure the comparison of local and global cohesion, which is widely used in group embeddedness~\cite{Krackhardt1988,Hanneman2011,Valentin2016}. We define the E-I index of GMNs as

\begin{equation}
\begin{split}
   \text{E-I index}_{(degree)}&=-\frac{EK-IK}{EK+IK}\\
\text{E-I index}_{(weight)}&=-\frac{EW-IW}{EW+IW} 
\end{split}
\end{equation}

The ``internal'' edge connects the two nodes in the same community, and the ``external'' edge connects the nodes from the different communities. $EK$ and $EW$ are the sums of external degrees and weights for all nodes, respectively; $IK$ and $IW$ are the sums of internal degrees and weights for all nodes, respectively. The E-I index ranges from 0 to 1. Smaller E-I index values indicate stronger connectivity between communities; larger E-I index values indicate stronger connectivity within the community and show that the community is more independent.

Fig. \ref{fig:EI} (d) shows a downward trend of the E-I index with respect to both degrees and weights. This figure indicates the continuous growth trend of cross-community connection and to some extent proves the significant trend of globalization in GMNs over the past fifty years.

\begin{figure}
    \centering
    \includegraphics[width=1.0\textwidth]{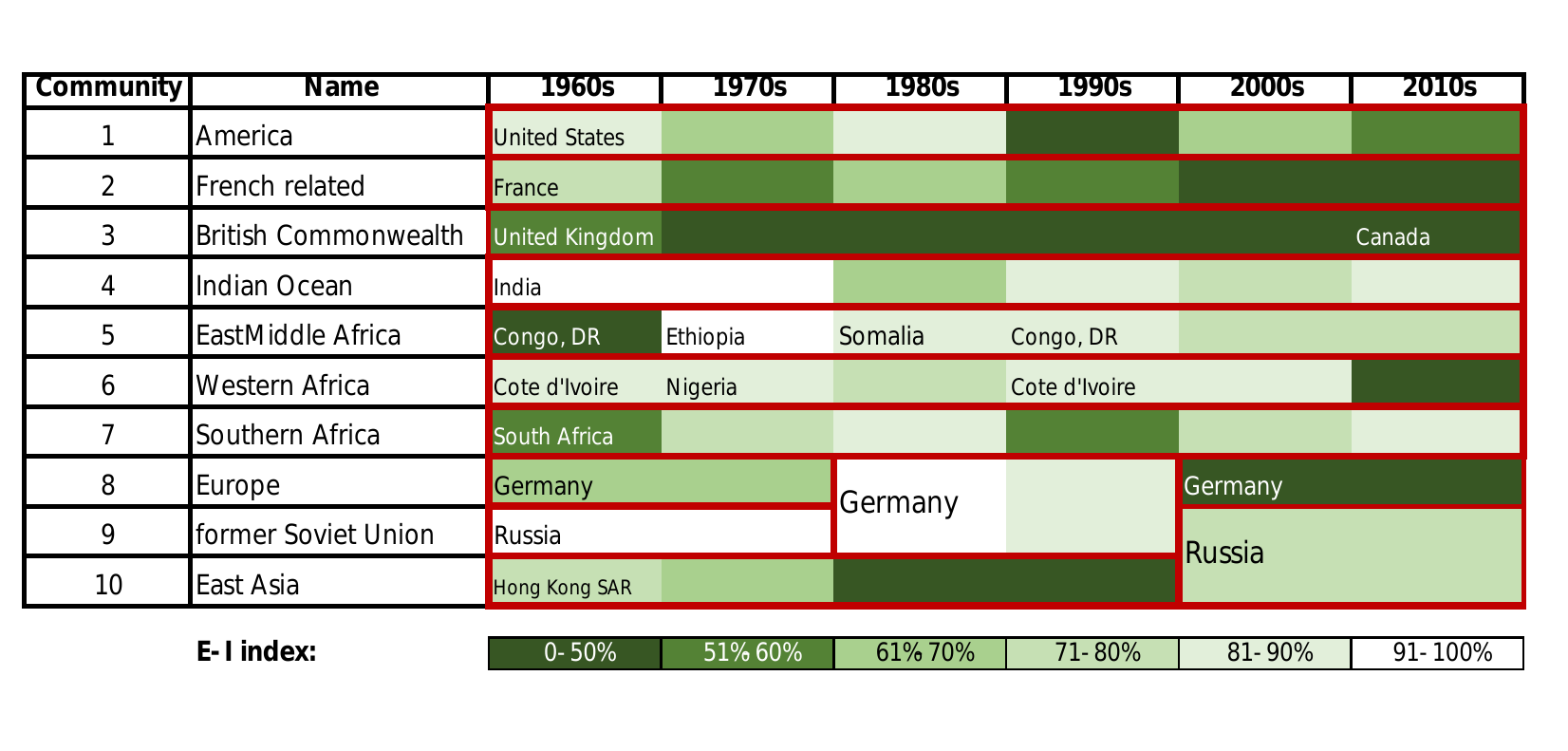}
    \caption{E-I index $_{(degree)}$ for each community from the 1960s to 2010s. The countries/regions having the largest degrees in the communities are listed in the table.}
    \label{fig:EI community}
\end{figure}

What's more, Fig. \ref{fig:EI community} shows E-I index values for ten typical communities; the countries/regions possessing the largest degree in the communities are listed, which could be regarded as the center of communities. For most communities, the central country/region is relatively stable and unchanged for these 50 years or is only adjusted between neighboring countries. It is worth noting that there are mergers and splits of communities 8-10, Europe, the former Soviet Union and East Asia, which will be described with details in Section \ref{section:structure evolution}. Here, dark green indicates the smaller E-I index values for the community in the corresponding time period.

The results present that over the past 50 years, the migration relation between different communities has become closer, which is reflected in the overall decline in E-I index values presented in Fig. \ref{fig:EI community}. In particular, the two communities of the former Soviet Union and Indian Ocean are typically introverted, with most of the migration flow coming from the ``internal edges'' connecting the community members; the communities of America, French related and Europe (since the 2000s) are extroverted, where the cross-community migration relation is greater than that within the communities. On the whole, the number of extroverted communities is increasing over time, and this also shows that for the potential immigrants, the possible moving routes among communities become more abundant. In contemporary times, the number of communities with E-I index values below 50\% (dark green grids) has increased from ONE to FOUR, which indicates that the GMNs became more globalized and multipolarized from the 1960s to 2010s.

\subsection{Structural evolution of GMNs}
\label{section:structure evolution}

\begin{figure}
    \centering
    \includegraphics[width=1.0\textwidth]{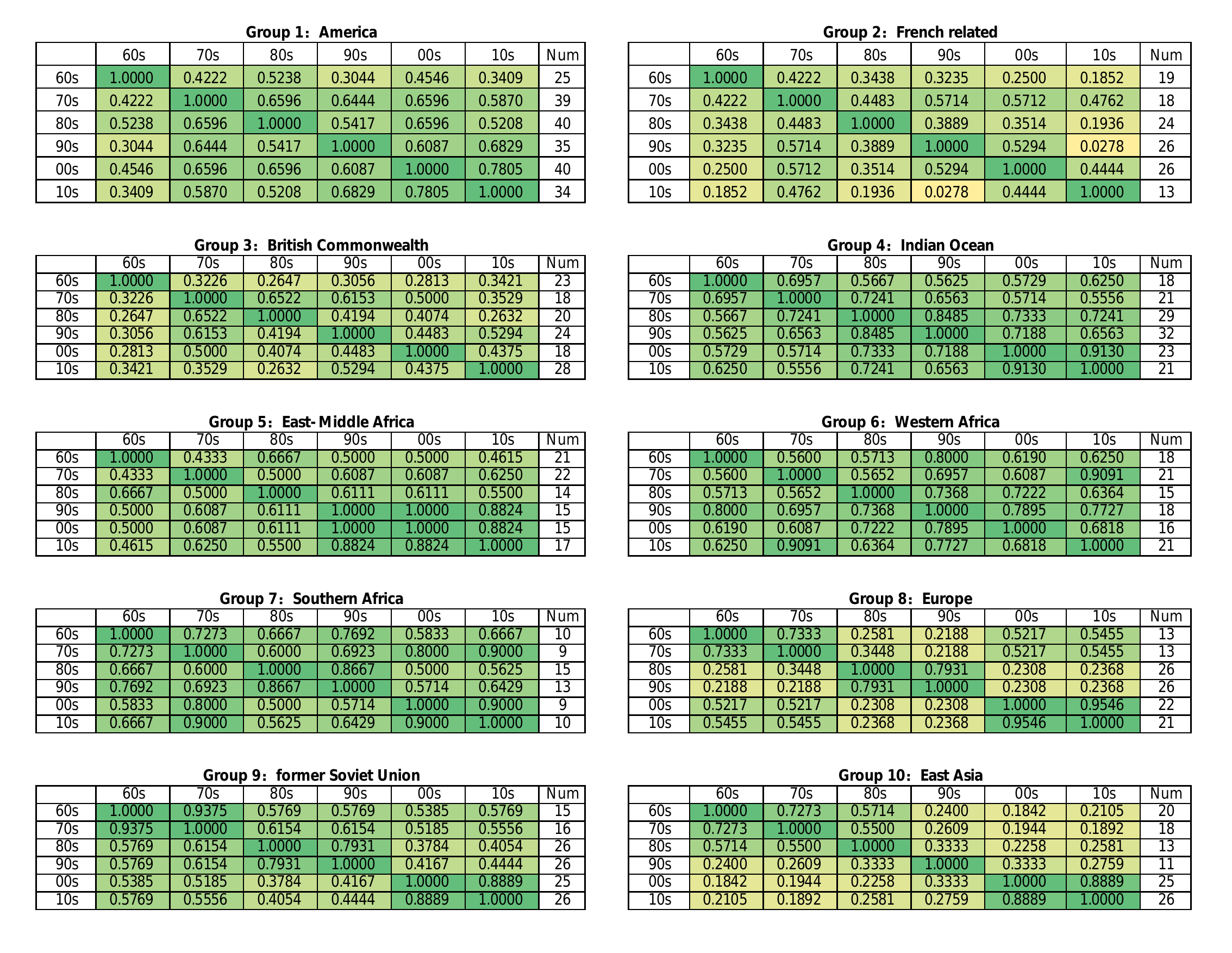}
    \caption{Jaccard similarity coefficients of ten typical communities for different time periods. ``Num'' means the number of members in the corresponding community.}
    \label{fig:clusters}
\end{figure}

\begin{figure}
    \centering
    \subfigure[1960-1969]{
    \includegraphics[width=0.9\textwidth]{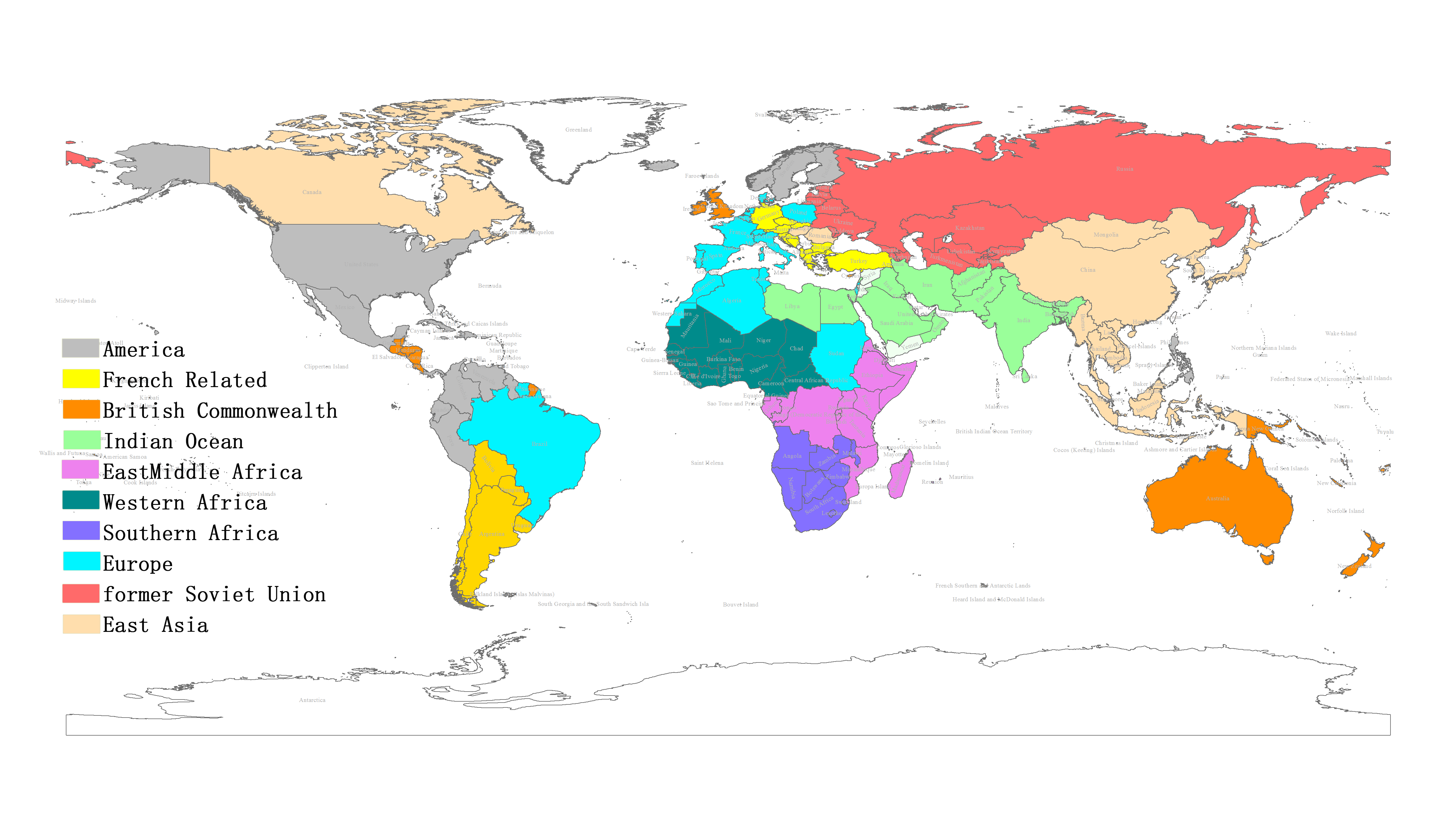}
}
    \subfigure[2010-2015]{
    \includegraphics[width=0.9\textwidth]{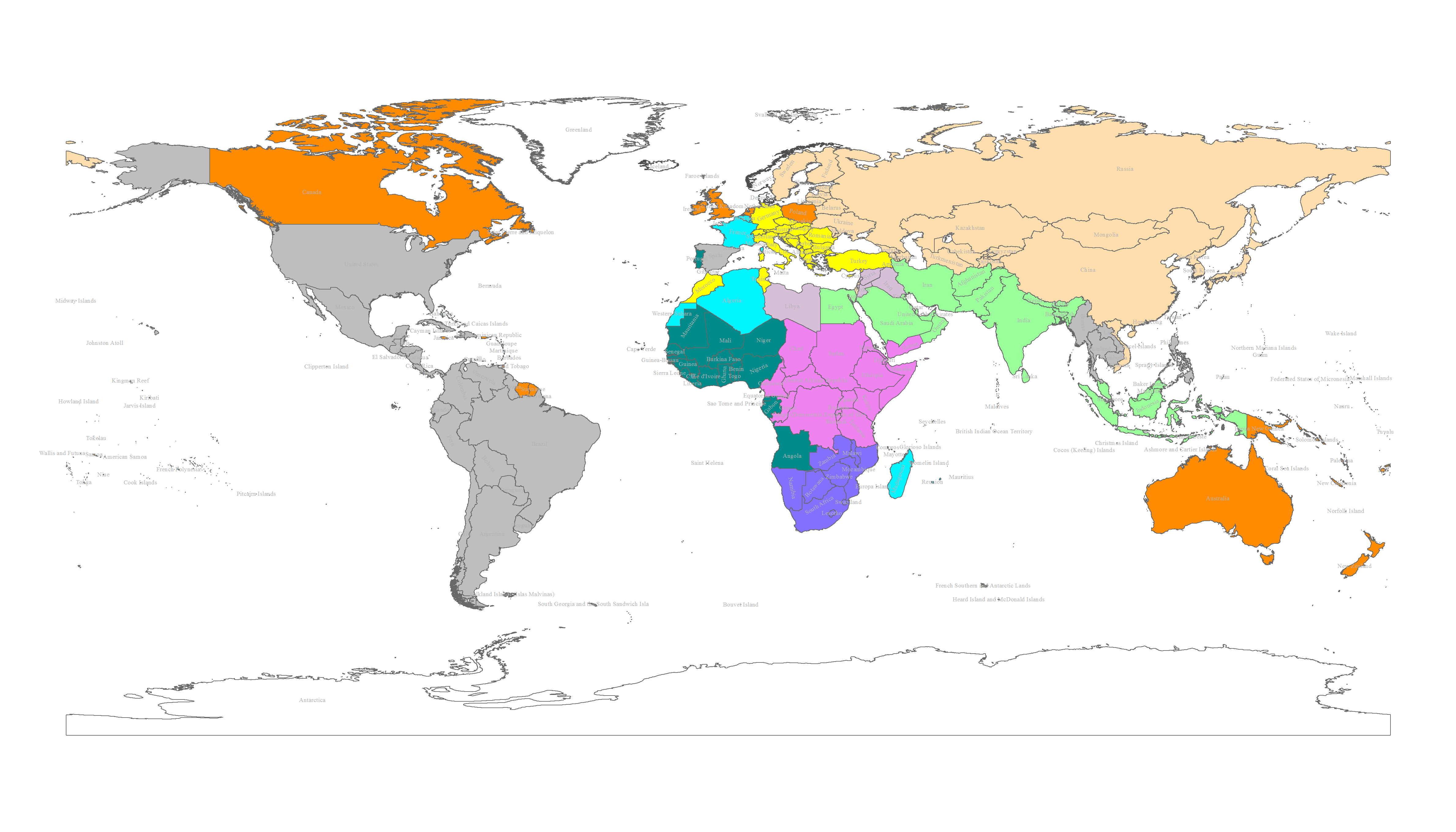}
}
    \caption{Community map of GMNs in the 1960s and 2010s showing the characteristics related to geographical distribution.}
    \label{fig:community-map}
\end{figure}

To assess the network structure more clearly, we analyze the correlation of the network communities with time. Fig. \ref{fig:clusters} shows the matrix of Jaccard similarity coefficients of ten typical communities covering 90-97\% of the countries/regions. In general, the composition of the members of each community is stable, and the characteristics related to geographical location are shown (Fig. \ref{fig:community-map}). Over more than 50 years, the central countries/regions of most communities have not changed. The green color indicates greater correlation, along with the higher coincidence of the members for the cluster between two eras. In contrast, the yellow color indicates that the structure of the communities changed greatly during this time.

Focusing on the yellow grids in Fig. \ref{fig:clusters}, combined with the specific composition of each community in Figs. \ref{fig:cluster} and \ref{fig:community-map}, we found some structural evolution of global migration networks during the past 50 years.

{\bfseries The community of former Soviet Union:} the ninth community, centered on Russia, was an independent cluster in the 1960s-1970s; in the 1980s-1990s, it merged into the community of Europe centered on Germany. Such structural changes may be related to the collapse of the former Soviet Union in 1991, when it pursued the policy of deporting the nonnative population, together with the boom of immigrants from the East into Western Europe~\cite{Krassinets2001Potentials,Kopnina2005East}. After 2000, the former Soviet Union cluster left the Germany group and merged into the community of East Asia. In fact, since then, Russia gradually replaced Hong Kong SAR as the new center of the community. The map also shows that Russia and these former Soviet Union countries have had a closer relationship with East Asia in GMNs since 2000 (Figure \ref{fig:community-map}).

{\bfseries The country of Canada:} once belonging to the British Commonwealth, Canada had a close relationship with Hong Kong SAR, and they were in the same community in the 1960s-1970s. Canada changed during 2000-2010 to join the community of America, which contained most of the American countries (Fig. \ref{fig:community-map}). In fact, some scholars have certified the relationships of the countries in Latin America and North America, including the United States~\cite{Durand2009Processes,Castillo2010A}. However, beginning in 2010, Canada left the United States community and became the new center of the British Commonwealth community; it is also the third closest country to the center on the hyperbolic plane, after the United States and French Guiana (Fig. \ref{fig:HyEm2015}).

{\bfseries The community centered on France:} the structure of the second community centered on France also substantially changed. In the 1990s, 68\% of its members belonged to European countries or their territories, but in the 2010s, the European members only accounted for 58\%. In contrast, in the 1960s, African countries accounted for only 15\% of this community, but after 2010, the proportion of African countries increased to 42\%, which also shows that France, as the representative and center of the community, became increasingly close to African countries in the global migration network. This trend should be closely related to the influence of language and historical colonies~\cite{Thomas2007Black,Li2016Characterizing}.

{\bfseries The countries including Malaysia, Singapore and Indonesia:} in the 1960s-1970s, Malaysia, Singapore and Indonesia were in the East Asia community, with Hong Kong SAR as the center. But since the 1980s, these three countries have been transferred to the community centered on India, which has greatly impacted the East Asia community and greatly reduced the number of its members. After 2000, the former Soviet Union community was merged and the center of the cluster was adjusted to Russia, which greatly changed the structure of the East Asia community again.

\section{Conclusion}

Global population migration is a typical complex system. At the micro level, each potential migrant makes a rational decision on ``whether'' and ``where'' to migrate according to the diversity utility function. Although the individuals are heterogeneous, specific migration patterns and evolution rules are continually emerging on the macro level.

The migration relationship between countries is complex and multilateral, and network theories could provide better description and more clearly exhibit on its structure and statistical characteristics. This paper constructs undirected global migration networks (GMNs) based on estimated bilateral migration flows during 1960-2015. The GMNs display the characteristics of disassortativity and high clustering with a typical power law in degree distribution. In the most recent fifty years, the network density and clustering have been increasing; the Gini coefficient of the degree and weight both exhibit an overall downward trend; and the entire network becomes more balanced and exhibits greater connectivity with time. 

From the network perspective, we analyze the evolution trend of international migration by comparing the global and local migration connectivity in communities. On the whole, the number of extroverted communities is increasing over time. This observation indicates the continuous growth trend of cross-community connection and, to some extent, proves the significant trends of ``globalization'' and ``multipolarization'' in the global migration network since the 1960s.

The existing literature does not discuss the geometric characteristics of the population migration network. This paper indicates that the GMNs exhibited a significant hyperbolic characteristic and hierarchical structure during 1960-2015, which is becoming more obvious these years. We embed the GMNs into hyperbolic space and finally obtain the locations of 200 countries/regions on a 2-dimensional Poincar\'e disk. 
Based on the definition of angular density distribution, it showed a trend of multi-centering of the global migration networks since 1990s.

Finally, we analyze the correlation between network communities and the structural evolution of GMNs with time. In general, from 1960 to 2015, the composition of the members of each community remained stable, and the central countries/regions of most communities did not change. In addition, we still find some changes: the former Soviet Union community merged into Germany during the 1980s-1990s, which could be related to the collapse of the former Soviet Union, and it left the German group and replaced Hong Kong SAR as the center of the East Asian community after 2000; the community centered in France reduced the proportion of members from Europe, and more African countries, especially those with colonial relations and labor contracts with France, gradually joined the group beginning in the 1960s; some southeastern Asian countries such as Singapore, Malaysia and Indonesia were in the East Asian community but transferred to the Indian Ocean community centered in India since the 1980s.

This paper provides a creative way to analyze the structural, statistical, and geometric characteristics and hierarchical structure of the population migration network. With respect to complex human migration behavior, it is far from sufficient to analyze only the migration flow data. In future research, we will consider the economic, social and policy factors that affect the decision-making of potential migrants and research the features and evolution of population migration patterns more comprehensively and scientifically.

\clearpage

\section*{Conflicts of Interest}
The authors declare no conflict of interest including any financial, personal or other relationships with other people or organizations.

\section*{Funding Statement}
This work was supported by the Chinese National Natural Science Foundation (71701018, 61673070); Humanities and Social Sciences Foundation of Ministry of Education of China (20YJAZH010); the National Social Sciences Fund, China (14BSH024); and the Beijing Normal University Cross-Discipline Project.

\section*{Acknowledgments}
We appreciate the comments and helpful suggestions from Professors Honggang Li, Handong Li and Yougui Wang. 

\clearpage

\bibliographystyle{splncs04}

\bibliography{migration}

\begin{thebibliography}{10}
\providecommand{\url}[1]{\texttt{#1}}
\providecommand{\urlprefix}{URL }
\providecommand{\doi}[1]{https://doi.org/#1}

\bibitem{GuyAbelDatabase}
Abel, G.: Bilateral international migration flow estimates for 200 countries.
  [EB/OL],
  \url{https://figshare.com/collections/Bilateral_international_migration_flow_estimates_for_200_countries/4470464}
  Accessed April 19, 2020

\bibitem{abel2018estimates}
Abel, G.J.: Estimates of global bilateral migration flows by gender between
  1960 and 20151. International Migration Review  \textbf{52}(3),  809--852
  (2018)

\bibitem{abel2019bilateral}
Abel, G.J., Cohen, J.E.: Bilateral international migration flow estimates for
  200 countries. Scientific data  \textbf{6}(1),  1--13 (2019)

\bibitem{abel2014quantifying}
Abel, G.J., Sander, N.: Quantifying global international migration flows.
  Science  \textbf{343}(6178),  1520--1522 (2014)

\bibitem{Barthelemy2000}
Amaral, L.A.N., Scala, A., Barthelemy, M., Stanley, H.E.: Classes of
  small-world networks. Proceedings of the national academy of sciences
  \textbf{97}(21),  11149--11152 (2000)

\bibitem{James2011The}
Anderson, J.E.: The gravity model. Annu. Rev. Econ.  \textbf{3}(1),  133--160
  (2011)

\bibitem{geonetcomp2015}
Asta, D., Shalizi, C.R.: Geometric network comparison. arXiv preprint
  arXiv:1411.1350  (2014)

\bibitem{azose2019estimation}
Azose, J.J., Raftery, A.E.: Estimation of emigration, return migration, and
  transit migration between all pairs of countries. Proceedings of the National
  Academy of Sciences  \textbf{116}(1),  116--122 (2019)

\bibitem{Bashan2013The}
Bashan, A., Berezin, Y., Buldyrev, S.V., Havlin, S.: The extreme vulnerability
  of interdependent spatially embedded networks. Nature Physics
  \textbf{9}(10),  667--672 (2013)

\bibitem{Beine2001Brain}
Beine, M., Docquier, F., Rapoport, H.: Brain drain and economic growth: theory
  and evidence. Journal of development economics  \textbf{64}(1),  275--289
  (2001)

\bibitem{BergstrandGravity2013}
Bergstrand, J.H., Egger, P., Larch, M.: Journal of International Economics
  \textbf{89}(1),  110--121 (2013)

\bibitem{louvain2008}
Blondel, V.D., Guillaume, J.L., Lambiotte, R., Lefebvre, E.: Fast unfolding of
  communities in large networks. Journal of statistical mechanics: theory and
  experiment  \textbf{2008}(10),  P10008 (2008)

\bibitem{boccaletti2006complex}
Boccaletti, S., Latora, V., Moreno, Y., Chavez, M., Hwang, D.: Complex
  networks: Structure and dynamics. Physics Reports  \textbf{424},  175--308
  (2006)

\bibitem{Michele2015Hyperbolicity}
Borassi, M., Chessa, A., Caldarelli, G.: Hyperbolicity measures democracy in
  real-world networks. Physical Review E  \textbf{92}(3),  032812 (2015)

\bibitem{Borjas1987Self}
Borjas, G.J.: Self-selection and the earnings of immigrants. Tech. rep.,
  National Bureau of Economic Research (1987)

\bibitem{Borjas1999Chapter}
Borjas, G.J.: The economic analysis of immigration. In: Handbook of labor
  economics, vol.~3, pp. 1697--1760. Elsevier (1999)

\bibitem{Burger2009gravity}
Burger, M., Van~Oort, F., Linders, G.J.: On the specification of the gravity
  model of trade: zeros, excess zeros and zero-inflated estimation. Spatial
  Economic Analysis  \textbf{4}(2),  167--190 (2009)

\bibitem{Caldarelli2007Scale}
Caldarelli, G.: Scale-free networks: complex webs in nature and technology.
  Oxford University Press (2007)

\bibitem{Castillo2010A}
Castillo, M.{\'A}.: A preliminary analysis of emigration determinants in
  mexico, central america, northern south america and the caribbean1.
  International Migration  \textbf{32}(2),  269--306 (1994)

\bibitem{errorfun2017}
Cvetkovski, A., Crovella, M.: Low-stress data embedding in the hyperbolic plane
  using multidimensional scaling. Appl. Math  \textbf{11}(1),  5--12 (2017)

\bibitem{Valentin2016}
Danchev, V., Porter, M.A.: Heterogeneity of global and local connectivity in
  spatial network structures of world migration. Available at SSRN 2755271
  (2016)

\bibitem{Ridolfi2013Global}
Davis, K.F., D'Odorico, P., Laio, F., Ridolfi, L.: Global spatio-temporal
  patterns in human migration: a complex network perspective. PloS one
  \textbf{8}(1) (2013)

\bibitem{Dennett2016Estimating}
Dennett, A.: Estimating an annual time series of global migration flows--an
  alternative methodology for using migrant stock data. Global Dynamics:
  Approaches from Complexity Science pp. 125--142 (2016)

\bibitem{educationalmigra2006}
Docquier, F., Marfouk, A.: International migration by education attainment,
  1990--2000. International migration, remittances and the brain drain pp.
  151--199 (2006)

\bibitem{Frederic2012}
Docquier, F., Rapoport, H.: Globalization, brain drain, and development.
  Journal of Economic Literature  \textbf{50}(3),  681--730 (2012)

\bibitem{Durand2009Processes}
Durand, J.: Processes of migration in latin america and the caribbean
  (1950-2008). Mpra Paper  \textbf{25}(S1),  S61--S67 (2009)

\bibitem{International-MigrationNetworke}
Fagiolo, G., Mastrorillo, M.: International migration network: Topology and
  modeling. Physical Review E  \textbf{88}(1),  012812 (2013)

\bibitem{PottsModel2004}
Fortunato, S., Latora, V., Marchiori, M.: Method to find community structures
  based on information centrality. Physical review E  \textbf{70}(5),  056104
  (2004)

\bibitem{C2009Diasporas}
Fr{\'e}d{\'e}ric~Docquier, Ozden~C., B.M., {\"O}zden, {\c C}.: Diasporas.
  Journal of Development Economics  \textbf{95}(1),  30--41 (2009)

\bibitem{Cell-type-specific2017}
Gal, E., London, M., Globerson, A., Ramaswamy, S., Reimann, M.W., Muller, E.,
  Markram, H., Segev, I.: Rich cell-type-specific network topology in
  neocortical microcircuitry. Nature neuroscience  \textbf{20}(7), ~1004 (2017)

\bibitem{SerranoThe2016}
Garc{\'\i}a-P{\'e}rez, G., Bogu{\~n}{\'a}, M., Allard, A., Serrano, M.{\'A}.:
  The hidden hyperbolic geometry of international trade: World trade atlas
  1870--2013. Scientific reports  \textbf{6},  33441 (2016)

\bibitem{HansonIncome2011}
Grogger, J., Hanson, G.H.: Income maximization and the selection and sorting of
  international migrants. Journal of Development Economics  \textbf{95}(1),
  42--57 (2011)

\bibitem{Hallwarddriemeier2017Trouble}
Hallward-Driemeier, M., Nayyar, G.: Trouble in the Making?: The Future of
  Manufacturing-led Development. World Bank Publications (2017)

\bibitem{Hanneman2011}
Hanneman, R.A., Riddle, M.: Concepts and measures for basic network analysis.
  The SAGE handbook of social network analysis pp. 340--369 (2011)

\bibitem{eumigra2015}
Herm, A., Poulain, M.: Economic crisis and international migration. what the eu
  data reveal? Revue europ{\'e}enne des migrations internationales
  \textbf{28}(4),  145--169 (2012)

\bibitem{Hirst2009Globalization}
Hirst, P., Thompson, G.: Globalization in question 3rd edition (cambridge.
  Polity  (2009)

\bibitem{Kopnina2005East}
Kopnina, H.: East to west migration: Russian migrants in western europe. Slavic
  Review  \textbf{66}(2),  371--372 (2005)

\bibitem{Krackhardt1988}
Krackhardt, D., Stern, R.N.: Informal networks and organizational crises: An
  experimental simulation. Social psychology quarterly pp. 123--140 (1988)

\bibitem{Krassinets2001Potentials}
Krassinets, E., Tiuriukanova, E.: Potentials of labour out-migration from
  russia: two surveys. Tijdschrift voor economische en sociale geografie
  \textbf{92}(1),  5--17 (2001)

\bibitem{BoguHyperbolic2010}
Krioukov, D., Papadopoulos, F., Kitsak, M., Vahdat, A., Bogun{\'a}, M.:
  Hyperbolic geometry of complex networks. Physical Review E  \textbf{82}(3),
  036106 (2010)

\bibitem{lancichinetti2010characterizing}
Lancichinetti, A., Kivel{\"a}, M., Saram{\"a}ki, J., Fortunato, S.:
  Characterizing the community structure of complex networks. PloS one
  \textbf{5}(8),  e11976 (2010)

\bibitem{LiAnalyzing2020}
Li, X., Huang, S., Chen, J., Chen, Q.: Analysis of the driving factors of us
  domestic population mobility. Physica A: Statistical Mechanics and its
  Applications  \textbf{539},  122984 (2020)

\bibitem{QinghuaAnalyzing2019}
Li, X., Huang, S., Chen, Q.: Analyzing the driving and dragging force in
  china’s inter-provincial migration flows. International Journal of Modern
  Physics C  \textbf{30}(07),  1940015 (2019)

\bibitem{Li2016Characterizing}
Li, X., Xu, H., Chen, J., Chen, Q., Zhang, J., Di, Z.: Characterizing the
  international migration barriers with a probabilistic multilateral migration
  model. Scientific reports  \textbf{6},  32522 (2016)

\bibitem{Galeano2011Castles}
Miller, M.J., Castles, S.: The age of migration: International population
  movements in the modern world. Palgrave Macmillan Basingstoke, Hampshire
  (2009)

\bibitem{GN2004}
Newman, M.E., Girvan, M.: Finding and evaluating community structure in
  networks. Physical review E  \textbf{69}(2),  026113 (2004)

\bibitem{M2003The}
Newman, M.: The structure and function of complex networks. Computer Physics
  Communications  \textbf{147}(1-2),  40--45 (2003)

\bibitem{Nickel2017Poincar}
Nickel, M., Kiela, D.: Poincar{\'e} embeddings for learning hierarchical
  representations. In: Advances in neural information processing systems. pp.
  6338--6347 (2017)

\bibitem{Peri2013The}
Ortega, F., Peri, G.: The effect of income and immigration policies on
  international migration. Migration Studies  \textbf{1}(1),  47--74 (2013)

\bibitem{RodrigoCommunity2016}
Peres, M., Xu, H., Wu, G.: Community evolution in international migration top1
  networks. PloS one  \textbf{11}(2) (2016)

\bibitem{multilayer2017}
Pilosof, S., Porter, M.A., Pascual, M., K{\'e}fi, S.: The multilayer nature of
  ecological networks. Nature Ecology \& Evolution  \textbf{1}(4), ~1--9 (2017)

\bibitem{PootGravity2016}
Poot, J., Alimi, O., Cameron, M.P., Mar{\'e}, D.C.: The gravity model of
  migration: the successful comeback of an ageing superstar in regional science
   (2016)

\bibitem{porat2016global}
Porat, I., Benguigui, L.: Global migration topology analysis and modeling of
  bilateral flow network 2006--2010. EPL (Europhysics Letters)
  \textbf{115}(1),  18002 (2016)

\bibitem{IdanPorat2015}
Porat, I., Benguigui, L.: World migration degree global migration flows in
  directed networks. arXiv preprint arXiv:1511.05338  (2015)

\bibitem{Ravenstein1884The}
Ravenstein, E.G.: The laws of migration. Journal of the statistical society of
  London  \textbf{48}(2),  167--235 (1885)

\bibitem{Roy1951}
Roy, A.D.: Some thoughts on the distribution of earnings. Oxford economic
  papers  \textbf{3}(2),  135--146 (1951)

\bibitem{Zolberg1989The}
Sassen, S.: The mobility of labor and capital: A study in international
  investment and labor flow. Cambridge University Press (1990)

\bibitem{Serrano2009}
Serrano, M.{\'A}., Bogun{\'a}, M., Vespignani, A.: Extracting the multiscale
  backbone of complex weighted networks. Proceedings of the national academy of
  sciences  \textbf{106}(16),  6483--6488 (2009)

\bibitem{Thomas2007Black}
Thomas, D.: Black France: Colonialism, immigration, and transnationalism.
  Indiana University Press (2006)

\bibitem{Nijkamp2014International}
Tranos, E., Gheasi, M., Nijkamp, P.: International migration: a global complex
  network. Environment and Planning B: Planning and Design  \textbf{42}(1),
  4--22 (2015)

\bibitem{un.org.migration}
the United~Nations: Un global migration database. [EB/OL],
  \url{http://www.un.org/en/development/desa/population/migration/data/}
  Accessed April 13, 2020

\bibitem{BeaucheminINTERNATIONAL2016}
Willekens, F., Massey, D., Raymer, J., Beauchemin, C.: International migration
  under the microscope. Science  \textbf{352}(6288),  897--899 (2016)

\bibitem{worldbank.org}
WorldBank: World bank global bilateral migration database. [EB/OL],
  \url{http://go.worldbank.org/092X1CHHD0} Accessed April 13, 2020

\bibitem{directem2020}
Wu, Z., Di, Z., Fan, Y.: An asymmetric popularity-similarity optimization
  method for embedding directed networks into hyperbolic space. Complexity
  \textbf{2020} (2020)

\bibitem{oiltrade}
Zhong, W., An, H., Gao, X., Sun, X.: The evolution of communities in the
  international oil trade network. Physica A: Statistical Mechanics and its
  Applications  \textbf{413},  42--52 (2014)

\bibitem{ZipfThe1946}
Zipf, G.K.: The p 1 p 2/d hypothesis: on the intercity movement of persons.
  American sociological review  \textbf{11}(6),  677--686 (1946)

\end{thebibliography}

\clearpage

\section*{Appendix}
\appendix
\renewcommand{\appendixname}{Appendix~\Alph{section}}

\section{Inhomogeneities and Disparity filter algorithm}
\label{appendix:Disparity filter algorithm}

To calculate inhomogeneities at the local level, for each country $i$ with $k$ migration routes, we calculate the Herfindahl-Hirschman index (HHI) $Y_i(k)$, which is extensively used as an economic standard indicator of market concentration, and it is also denoted as the disparity measure in the complex networks literature,

\begin{equation}
Y_i(k)=\sum_j \left(\frac{\omega_{i,j}}{s_i}\right)^2
\label{}
\end{equation}

where $\omega_{ij}$ is the total flow between countries $i$ and $j$ and $s_i = \sum_j \omega_{i,j} $ is the strength (aggregated migration) of country $i$. If country $i$ distributes its migration homogeneously between its migration partners, then $kY_i(k)= 1$; in the opposite case, if all its migration is concentrated on a single link, then $kY_i(k) = k$. For the inhomogeneities network, we can use the disparity filter to extract the backbone.

The disparity filter proceeds as follows. We first normalize the weights of edges linking node $i$ with its neighbor $j$ as $p_{i,j} = \omega_{i,j}/s_i$, with $s_i=\sum_j\omega_{i,j}$ being the strength of node $i$ and $\omega_{i,j}$ the weight of the edge connecting $i$ and $j$. For each migration channel of a given country $i$, we compute the probability $\alpha_{i,j}$ that the link takes the observed value $p_{i,j}$ according to the purely random null model. By imposing a significance level $\alpha$, we can determine the statistical significance of a given migration channel by comparing $\alpha_{i,j}$ to $\alpha$. Therefore, if $\alpha_{i,j}>\alpha$, the flow through that migration channel can be considered compatible with a random distribution (with the chosen significance level $\alpha$) and is thus discarded. The statistically relevant channels are those that satisfy

\begin{equation}
\alpha_{i,j}=1-(k-1)\int_{0}^{p_{i,j}} (1-x)^{k-2}\,dx<\alpha
\label{}
\end{equation}
for at least one of the two countries $i$ and $j$. $k$ represents the degree of node $i$.

By applying this selection rule to all of the links in the network, we find the backbone, a new graph containing, in general, fewer links and nodes, as the GMN in this paper. However, the number of links and nodes removed depends on the value of the significance level $\alpha$. To find the appropriate value of $\alpha$, it is convenient to plot the fraction of remaining nodes $N_{BB}/N$ and the fraction of remaining weights $W_{BB}/W$ in the backbone vs. the fraction of remaining links $L_{BB}/L$ for different values of $\alpha$. As the filter becomes more restrictive, the number of links decreases while keeping almost all nodes until a certain critical point, after which the number of nodes begins a steep decay. To retain more countries, more weights and fewer links, we choose a point where the number of nodes begins to be lower than the initial value as our specific indicator $\alpha_s$ for extracting the backbone networks~\cite{Serrano2009}.

\section{Hyperbolic embedding method and evaluation index}
\label{appendix:effectiveness of hyperbolic embedding}

The method proposed by Maximilian Nickel and Douwe Kiela is based on the Poincaré ball model, as it is well suited for gradient-based optimization~\cite{Nickel2017Poincar}. In particular, let $\mathcal{B}^d = {x \in \mathbb{R}^d | \lVert x \rVert < 1}$ be the \textit{open d}-dimensional unit ball, where $\lVert \cdot \rVert$ denotes the Euclidean norm. The Poincaré ball model of hyperbolic space then corresponds to the Riemannian manifold ($\mathcal{B}^d$, $g_{\mathbf{x}}$), i.e., the open unit ball equipped with the Riemannian metric tensor

\begin{equation}
g_{\mathbf{x}}=\left(\frac{2}{1-{ \lVert x \rVert}^2}\right)^2g^{\textit{E}}
\label{}
\end{equation}
where $x \in \mathcal{B}^d$ and $g^{\textit{E}}$ denotes the Euclidean metric tensor. Furthermore, the distance between points $\mathbf{u,v} \in\mathcal{B}^d$ is given as

\begin{equation}
d(\mathbf{u,v})=arcosh\left(1+2\frac{{\lVert\mathbf{u-v}\rVert}^2}{(1-{\lVert\mathbf{u}\rVert}^2)(1-{\lVert\mathbf{v}\rVert}^2)}\right)
\label{duv}
\end{equation}

Note that Equation \ref{duv} is symmetric and that the hierarchical organization of the space is solely determined by the distance of nodes to the origin. Due to this self-organizing property, Equation \ref{duv} is applicable in an unsupervised setting where the hierarchical order of objects such as text and networks is not specified in advance. Remarkably, Equation \ref{duv} therefore allows us to learn embeddings that simultaneously capture the hierarchy of objects (through their norms) as well as their similarity.

We embedded our GMN in all time periods (1960-2015), using a 2-dimensional Poincaré disk, with a learning rate of 0.1 and a negative sample size of 30. To further explain the embedding effect, the general form of the function is as below:

\begin{equation}
E=c\sum_{j=1}^n\sum_{k=j+1}^n c_{j,k}(d_{j,k}-a\delta_{j,k})^2 
\label{eq:error}
\end{equation}
where $\delta_{j,k}$ is the dissimilarity between nodes $j$ and $k$ and $d_{j,k}$ represents the embedded distances. Eq.(\ref{eq:error}) is a general form from which several special embedding error functions can be obtained by substituting
appropriate values of the constants $c$, $c_{jk}$, and $a$\cite{errorfun2017}. In our calculate, we made all of the constants equal to 1 for simplify. To transfer the migration matrix to the dissimilarity matrix, for every weight $\omega_{i,j}$, we use $\sqrt{1-(\omega_{i,j} / \omega_{max})}$ ($\omega_{max}$ denotes the maximum weight in the matrix) to replace the original data. In Euclidean space, we use two kinds of regular MDS (multidimensional scaling) methods, namely, the nonmetric MDS (hereinafter referred to as NMM) and nonclassical MDS (hereinafter referred to as NCM), to embed the data for comparison. 

The error function described in the previous section calculates the cumulative difference between the embedded distance and the actual data. However, it also concerns another issue: whether the two countries with closer relations are actually closer to each other than other countries after embedding. Here, we propose a scoring scheme to assess this possibility. For any two edges $l_{i,j}$ and $l_{m,n}$ that exist in the network, where $i\ne m$ and $j\ne n$, and with corresponding embedding distances $d_{i,j}$ and $d_{m,n}$, we calculate that

\begin{equation}
S=\begin{cases}
1 \quad (l_{i,j}-l_{m,n})(d_{i,j}-d_{m,n})<0\\
0 \quad otherwise
\end{cases}
\label{}
\end{equation}

We repeat this random selection n times and obtain the following scores $Score=100\cdot\frac{\sum S}{N}$. Since there are almost no consistent original migration data or embedding distances between different countries, $(l_{i,j}-l_{m,n})(d_{i,j}-d_{m,n})=0$ is almost nonexistent. Thus, $Score$ indicates the probability of meeting the required rules for any two links. The closer the $Score$ is to 100, the better the embedding distance can interpret the size relationship in the original data. 

\section{Statistical characteristics of GMN}
\label{appendix:Statistical characteristics of GMN}

Table \ref{tab:overall} shows the statistical characteristics and evolution trends of global migration networks (Note that: APL: Average path length; CC: Clustering coefficient; ND: Node degree; NS: Node strength). We can observe an increase in the edges and weights, along with increased network density and connectivity, which means that many countries have become closer to each other in the GMNs.

\renewcommand{\arraystretch}{1.5} 
\begin{table}[H]
 
  \begin{center}
  \fontsize{10}{7}\selectfont
  \caption{Descriptive statistics regarding the GMN.}
  \label{tab:overall}
    \begin{tabular}{ccccccc}
    \toprule
     &1960-1969&1970-1979&1980-1989&1990-1999&2000-2009&2010-2015 \cr
    \midrule
   Number of Node & 195 & 195&196&200&200&200  \cr
 Number of Edge  & 269 & 352&297&450&418&385 \cr
 Number of Community & 13 & 10&9&10&10&11  \cr
 APL  & 4.96 & 4.00 & 4.90 &3.54&3.86&4.11  \cr
 CC  & 0.18 & 0.26 & 0.27 & 0.45 & 0.43 & 0.34\cr
 Mean ND & 2.76& 3.61 &3.03&4.41&4.07&3.79\cr
 Max. ND  &27 &58 & 45&66&52&52\cr
 Std. ND  & 3.05 & 4.87&3.93&5.58&4.81&4.73\cr
 Mean NS  & 2.34E+5 &3.44E+6&3.49E+5&9.47E+5&1.01E+6&1.14E+6\cr
 Max. NS  & 4.85E+6&7.48E+6&7.90E+6&2.36E+7&2.57E+7&2.63E+7\cr
 Std. NS  & 5.23E+5&8.69E+5&8.55E+5&2.22E+6&2.34E+6&2.52E+6\cr
    \bottomrule
    \end{tabular}
    \end{center}
\end{table}

\section{Results of community division}
We use the Louvain algorithm to divide the community of all networks. For example, the complete community results from 2010 to 2015 are presented in Table \ref{tab:community}.

\begin{table}[tp]
 
  \centering
  \fontsize{10}{7}\selectfont
  \caption{Community of GMN in 2010-2015.}
  \label{tab:community}
    \begin{tabular}{ccp{8cm}}
    \toprule
     Group&Name& \makebox[12cm][c]{Members} \cr
    \midrule
    1&America&
    Aruba, Argentina, Antigua and Barbuda, Bahamas, Belize, Bolivia, Brazil, Chile, Colombia, Costa Rica, Cuba, Dominican Republic, Ecuador, Spain, Micronesia, Fed. Sts., Guatemala, Guam, Guyana, Honduras, Haiti, Cambodia, Lao PDR, Mexico, Myanmar, Nicaragua, Panama, Peru, Philippines, Paraguay, El Salvador, Thailand Uruguay, United States, Venezuela, RB.\cr
    \hline
    2&French related&
    Belgium, Comoros, Algeria, Western Sahara, France, Guadeloupe, Luxembourg, Madagascar, Martinique, Mauritius, Mayotte, French Polynesia, Réunion.\cr
    \hline
    3&British Commonwealth &
    Australia, Barbados, Canada, Channel Islands, Curaçao, Fiji, United Kingdom, Grenada, French Guiana, Ireland, Jamaica, Kiribati, St. Lucia, Malta, New Caledonia, Netherlands, New Zealand, Papua New Guinea, Poland, Puerto Rico, Solomon Islands, Suriname, Tonga, Trinidad and Tobago, St. Vincent and the Grenadine, Virgin Islands (U.S.), Vanuatu, Samoa.\cr
    \hline
    4&Indian Ocean&
    Afghanistan, United Arab Emirates, Bangladesh, Bahrain, Brunei Darussalam, Bhutan, Egypt, Arab Rep., Indonesia, India, Iran, Islamic Rep., Kuwait, Sri Lanka, Maldives, Malaysia, Nepal, Oman, Pakistan, Qatar, Saudi Arabia, Singapore, Timor-Leste.\cr
    \hline
    5&EastMiddle Africa&
    Burundi, Central African Republic, Cameroon, Congo, Dem. Rep., Congo, Rep., Djibouti, Eritrea, Ethiopia, Kenya, Rwanda, Sudan, Somalia, South Sudan, Chad, Tanzania, Uganda, Yemen, Rep..\cr
    \hline
    6&Western Africa&
    Angola, Benin, Burkina Faso, Cote d'Ivoire, Cabo Verde, Gabon, Ghana, Guinea, Gambia, The, Guinea-Bissau, Equatorial Guinea, Liberia, Mali, Mauritania, Niger, Nigeria, Portugal, Senegal, Sierra Leone, Sao Tome and Principe, Togo.\cr
    \hline
    7&Southern Africa&
    Botswana, Lesotho, Mozambique, Malawi, Namibia, Swaziland, Seychelles, South Africa, Zambia, Zimbabwe.\cr
    \hline
    8&Europe &
    Albania, Austria, Bulgaria, Bosnia and Herzegovina, Switzerland, Cyprus, Czech Republic, Germany, Greece, Croatia, Hungary, Italy, Morocco, Macedonia, FYR, Montenegro, Romania, Serbia, Slovak Republic, Slovenia, Tunisia, Turkey.\cr
    \hline
    9&former Soviet Union &
    Armenia, Azerbaijan, Belarus, China, Estonia, Finland, Georgia, Hong Kong SAR, China, Israel, Japan, Kazakhstan, Kyrgyz Republic, Korea, Rep., Lithuania, Latvia, Macao SAR, China, Moldova, Mongolia, Korea, Dem. Rep., Russian Federation, Sweden, Tajikistan, Turkmenistan, Ukraine, Uzbekistan, Vietnam.\cr
  \bottomrule
    \end{tabular}
\end{table}

\end{document}